\documentclass[11t, a4paper]{article}
\pdfoutput=1
\usepackage{jcappub}
\usepackage[utf8]{inputenc}
\usepackage[english]{babel}
\usepackage{amsmath, amssymb}
\usepackage{graphicx}
\usepackage{hyperref}

\newcommand*{\e}{\mathrm{e}}

\makeatletter
\newcommand*{\dd}{\mathop{}\!{\operator@font d}}
\makeatother

\newcommand*{\mH}{\mathcal{H}}
\newcommand*{\vk}{\mathbf{k}}
\newcommand*{\vq}{\mathbf{q}}

\newcommand*{\vv}{\mathbf{v}}

\newcommand*{\zmatch}{z_{\mathrm{match}}}
\newcommand*{\etamatch}{\eta_{\mathrm{match}}}

\newcommand*{\etaini}{\eta_{\mathrm{ini}}}

\newcommand*{\Mnu}{M_{\nu}}
\newcommand*{\fnu}{f_{\nu}}
\newcommand*{\cseff}{c_{s,\mathrm{eff}}}
\newcommand*{\sigmad}{\sigma_{\mathrm{d}}^2}

\newcommand*{\Pw}{P_{\mathrm{w}}}
\newcommand*{\Pnw}{P_{\mathrm{nw}}}

\newcommand*{\kosc}{k_{\mathrm{osc}}}
\newcommand*{\kpivot}{k_{\mathrm{pivot}}}
\newcommand*{\kmax}{k_{\mathrm{max}}}
\newcommand*{\kmin}{k_{\mathrm{min}}}
\newcommand*{\kFS}{k_{\mathrm{FS}}}
\newcommand*{\kNL}{k_{\mathrm{NL}}}

\newcommand*{\cb}{\mathrm{cb}}
\newcommand*{\cdm}{\mathrm{cdm}}
\newcommand*{\mm}{\mathrm{m},\mathrm{m}}

\newcommand*{\IR}{\mathrm{IR}}

\newcommand*{\oneloop}{\text{1-loop}}
\newcommand*{\twoloop}{\text{2-loop}}

\newcommand*{\ihMpc}{h\ \text{Mpc}^{-1}}
\newcommand*{\Mpc}{\mathrm{Mpc}}
\newcommand*{\eV}{\mathrm{eV}}

\newcommand*{\LCDM}{$\Lambda$CDM}

\newcommand*{\bea}{\begin{eqnarray}}
\newcommand*{\eea}{\end{eqnarray}}
\newcommand*{\be}{\begin{equation}}
\newcommand*{\ee}{\end{equation}}
\newcommand*{\nn}{\nonumber}

\subheader{\small%\sc
    \vspace*{-1.28cm}
    \begin{flushright}
        TUM-HEP-1402/22
    \end{flushright}
}

\title{Two-loop power spectrum with full time- and scale-dependence and EFT corrections: impact of massive neutrinos and going beyond EdS}

\author{Mathias Garny,}
\author{Petter Taule}

\affiliation{Physik Department T31,\\
    James-Franck-Str.\ 1, Technische Universit\"at M\"unchen,\\
D-85748 Garching, Germany}

\emailAdd{mathias.garny@tum.de}
\emailAdd{petter.taule@tum.de}

\abstract{%
We compute the density and velocity power spectra at next-to-next-to-leading order taking
into account the effect of time- and scale-dependent growth of massive neutrino perturbations as well as the
departure from Einstein--de-Sitter (EdS) dynamics at late times non-linearly. We determine
the impact of these effects by comparing to the commonly adopted approximate treatment where
they are not included. For the bare cold dark matter (CDM)+baryon spectrum, we find percent
deviations for $k\gtrsim 0.17\ihMpc$, mainly due to the departure from EdS. For
the velocity and cross power spectrum the main difference arises due to time- and scale-dependence in presence of massive neutrinos
yielding percent deviation above $k\simeq 0.08, 0.13, 0.16\ihMpc$ for $\sum
m_{\nu} = 0.4, 0.2, 0.1~\eV$, respectively. We use an effective field theory (EFT) framework at
two-loop valid for wavenumbers $k \gg \kFS$, where $\kFS$ is the neutrino free-streaming scale. Comparing
to Quijote N-body simulations, we find that for the CDM+baryon density
power spectrum the effect of neutrino perturbations and exact time-dependent dynamics at late times can be accounted
for by a shift in the one-loop EFT counterterm, $\Delta\bar{\gamma}_1 \simeq - 0.2~\Mpc^2/h^2$.
We find percent agreement between the perturbative and N-body results up to
$k\lesssim 0.12\ihMpc$ and $k\lesssim 0.16\ihMpc$ at one- and two-loop order, respectively, for all
considered neutrino masses $\sum m_{\nu} \leq 0.4~\eV$.
}

\begin{document}
\maketitle

\section{Introduction}

Ongoing and upcoming large-scale structure (LSS) surveys are expected to return
a rich amount of information that likely will yield valuable insights into the
properties of the dark components of the Universe, gravity on large scales as
well as the initial conditions from the early
Universe~\cite{SPT:2018njh, DES:2017myr, eBOSS:2020yzd, PFSTeam:2012fqu,
EuclidTheoryWorkingGroup:2012gxx, LSSTScience:2009jmu, DESI:2019jxc, 2021MNRAS.tmp.1608E}.
Moreover, they offer the prospect of measuring the absolute neutrino mass
scale~\cite{Lesgourgues:2006nd, Audren:2012vy, Amendola:2016saw, Boyle:2017lzt,
Mishra-Sharma:2018ykh,Sprenger:2018tdb, Brinckmann:2018owf, Chudaykin:2019ock, Hahn:2019zob,
Xu:2020fyg, Boyle:2020rxq}. Given the extended coverage
of larger scales and increased redshift of future probes, it is of major
importance to model structure formation in the mildly non-linear regime
accurately. Perturbative approaches~\cite{Bernardeau:2001qr, Crocce:2005xy} have gained a lot of attention, and
over the last decades been improved by including elements such as effective
field theory (EFT) methods~\cite{Baumann:2010tm,Carrasco:2012cv}, bias
expansion~\cite{Desjacques:2016bnm}, IR resummation~\cite{Eisenstein:2006nj,Seo:2007ns,Senatore:2014via,
Baldauf:2015xfa, Vlah:2015zda, Blas:2016sfa}, redshift-space distortions
(RSD)~\cite{Scoccimarro:1999ed, Scoccimarro:2004tg, Vlah:2012ni, Villaescusa-Navarro:2017mfx}, as well as techniques for fast
evaluation~\cite{McEwen:2016fjn, Schmittfull:2016jsw, Simonovic:2017mhp}. The
matter power spectrum has been computed up to three-loop in the
EFT~\cite{Konstandin:2019bay} and the bispectrum up to
two-loop~\cite{Baldauf:2021zlt}. In particular, perturbation theory has been
applied to analyse the full shape BOSS galaxy clustering data at the level of
the one-loop power spectrum~\cite{Montesano:2011bp, Sanchez:2013uxa, Sanchez:2013tga, DAmico:2019fhj, Ivanov:2019pdj,
Troster:2019ean, Semenaite:2021aen} (see also~\cite{Chen:2021wdi}) as well as tree-level and
one-loop bispectrum~\cite{Gil-Marin:2016wya,Philcox:2021kcw,DAmico:2022gki,Cabass:2022ymb} (see also~\cite{Pezzotta:2021vfn,Eggemeier:2021cam}).

The sum of neutrino masses is known to be greater than $0.06~\eV$ from
oscillation experiments~\cite{Esteban:2020cvm}, and $\beta$-decay experiments
sets an upper bound $m_{\beta} < 0.8~\eV$ at 90\% C.L. for the effective electron
anti-neutrino mass~\cite{KATRIN:2021uub}. Cosmological probes offer
complementary constraints because of observable impacts left by the neutrinos on
the cosmological evolution. In particular, mainly due to late ISW and lensing
effects, CMB measurements constrain the sum of neutrino masses to
$\Mnu \equiv \sum m_{\nu} \leq 0.26~\eV$ at 95\% C.L.~\cite{Planck:2018vyg}, improving to $\Mnu \leq 0.12~\eV$ when combined with
data on the scale of baryon acoustic oscillations (BAO). In addition,
the large neutrino velocity dispersion after their non-relativistic transition
in the late Universe slows down structure formation on scales smaller than the
neutrino free-streaming length, corresponding to a wavenumber $\kFS \sim 0.01\ihMpc$~\cite{Lesgourgues:2006nd}.
This characteristic impact is expected to be detectable by the Euclid
satellite, yielding a forecasted measurement of the neutrino mass sum with
$\sigma(\Mnu) \simeq 0.02$--$0.03~\eV$~\cite{Lesgourgues:2006nd, Audren:2012vy, Amendola:2016saw, Boyle:2017lzt,
Mishra-Sharma:2018ykh,Sprenger:2018tdb, Brinckmann:2018owf, Chudaykin:2019ock, Hahn:2019zob,
Xu:2020fyg, Boyle:2020rxq}. On the other hand, the presence of the neutrino free-streaming
scale introduces a scale-dependence in the clustering dynamics, making the
modelling of neutrino perturbations beyond linear theory complicated.

In order to model the mildly non-linear scales in a robust and efficient way
as necessary for MCMC analyses, certain approximations have to be made.
Accordingly, the approximations should be scrutinized both in $\Lambda$CDM and
extensions of it to determine the theoretical uncertainty. In this work we
perform a precision calculation of the matter power spectrum with the aim of
examining the accuracy of certain common approximations. In particular, our
analysis involve the following:
\begin{itemize}
    \item computing the matter density and velocity power spectrum at
        \emph{next-to-next-to-leading} order (NNLO) in perturbation theory,
        using the method that can capture scale- and time-dependent dynamics
        introduced in~\cite{Garny:2020ilv} (see also~\cite{Blas:2013bpa,
        Blas:2013aba, Blas:2015tla}),
    \item including neutrino perturbations up to fifth order (i.e.\ at
        two-loop) and their impact on the cold dark matter (CDM) and baryons fluid
        via gravity, utilizing a hybrid two-component fluid scheme introduced
        in~\cite{Blas:2014} and further developed in~\cite{Garny:2020ilv}\footnote{See also~\cite{Zennaro:2016nqo,Kamalinejad:2022yyl} for applications of a two-fluid setup at linear order
        as well as \cite{Fuhrer:2014zka,Chen:2020bdf} for attempts to include higher moments of the neutrino distribution non-linearly.},
    \item relaxing the Einstein--de-Sitter (EdS)
        approximation~\cite{Takahashi:2008yk,Fasiello:2016qpn,Garny:2020ilv,Donath:2020abv,Steele:2020tak,Baldauf:2021zlt,Fasiello:2022lff} and taking into account the
        exact $\Lambda$CDM ($+\Mnu$) time-dependence at two-loop order (as first done in~\cite{Garny:2020ilv}, see also~\cite{Fasiello:2022lff}),
    \item extending the work of~\cite{Garny:2020ilv} by including EFT
        corrections to the power spectrum in the two-component fluid model,
        promoting the approach of~\cite{Baldauf:2015aha} to a cosmology with
        scale-dependent dynamics, and further resumming the effect of large bulk flows in the
        IR~\cite{Eisenstein:2006nj,Seo:2007ns,Senatore:2014via,
Baldauf:2015xfa, Vlah:2015zda, Blas:2016sfa},
    \item comparing theory predictions to and calibrating EFT parameters using
        N-body data from the Quijote simulation
        suite~\cite{Villaescusa-Navarro:2019bje} for three cosmologies with
        $\Mnu = 0.1$, $0.2$ and $0.4~\eV$.
\end{itemize}
Our work extends previous studies~\cite{Senatore:2017hyk, deBelsunce:2018xtd, Chudaykin:2019ock, Aviles:2021que} for massive neutrinos within the EFT context in evaluating the power spectrum
at NNLO (two-loop) order, and including the full scale- and time-dependence of neutrinos imprinted on non-linear kernels.
The EFT setup used in this work is based on~\cite{Baldauf:2015aha}, and we argue that it can be extended to the case of
massive neutrinos in the limit $k\gg \kFS \simeq 0.01\ihMpc$, being satisfied for current and future galaxy surveys. Nevertheless,
we stress that our unrenormalized results capture also the scale dependence in the transition region $k\sim \kFS$.

With these elements, we can perform an assessment of the accuracy of neglecting
neutrino perturbations as well as the EdS approximation on the power spectrum
at two-loop order:
\begin{itemize}
\item We first compare the unrenormalized CDM+baryon power spectrum
between the full solution including non-linear neutrino perturbations and
departure from EdS (which we name the 2F scheme) to the simplified treatment
with linear neutrinos and EdS dynamics for CDM+baryons (named 1F scheme).
We find deviations beyond one percent for $k\gtrsim 0.17\ihMpc$ at $z=0$,
mainly arising from the departure from EdS.
\item As a proxy for redshift space distortion effects, we analyze the neutrino mass dependence of the
velocity divergence power spectrum and the cross spectrum. We indeed find deviations between the full 2F and approximate 1F scheme that strongly depend on the value
of the neutrino mass, and reach more than one percent for $k\gtrsim 0.08, 0.13, 0.16 \ihMpc$ for either the cross- or velocity spectrum and $M_\nu=0.4, 0.2, 0.1\,\eV$, respectively.
Therefore, the 1F approximation is not sufficient to access the neutrino mass information encoded in redshift space~\cite{Aviles:2021que, Bayer:2021kwg}.
\item After including EFT corrections, we
compare both 2F and 1F results to N-body data, and determine to what extent the
discrepancy of the simplified treatment can be absorbed by a readjustment of counterterms at NNLO. For the CDM+baryon density spectrum we find that this is indeed the case
to high accuracy, with a shift mainly in the one-loop EFT term by $\Delta\bar{\gamma}_1 \simeq - 0.2~\Mpc^2/h^2$.
\end{itemize}
We also address the level of degeneracy between neutrino free-streaming suppression and the overall amplitude of fluctuations~\cite{Bayer:2021kwg} for the power spectrum within perturbation theory.
We note that the approach presented here can be extended to the (one-loop) bispectrum in a straightforward manner. It has been shown that this additional information
could be instrumental to break the degeneracy of neutrino masses with the overall power spectrum amplitude~\cite{Hahn:2019zob,Hahn:2020lou}.

\begin{figure}
    \centering
    \includegraphics[width=\textwidth]{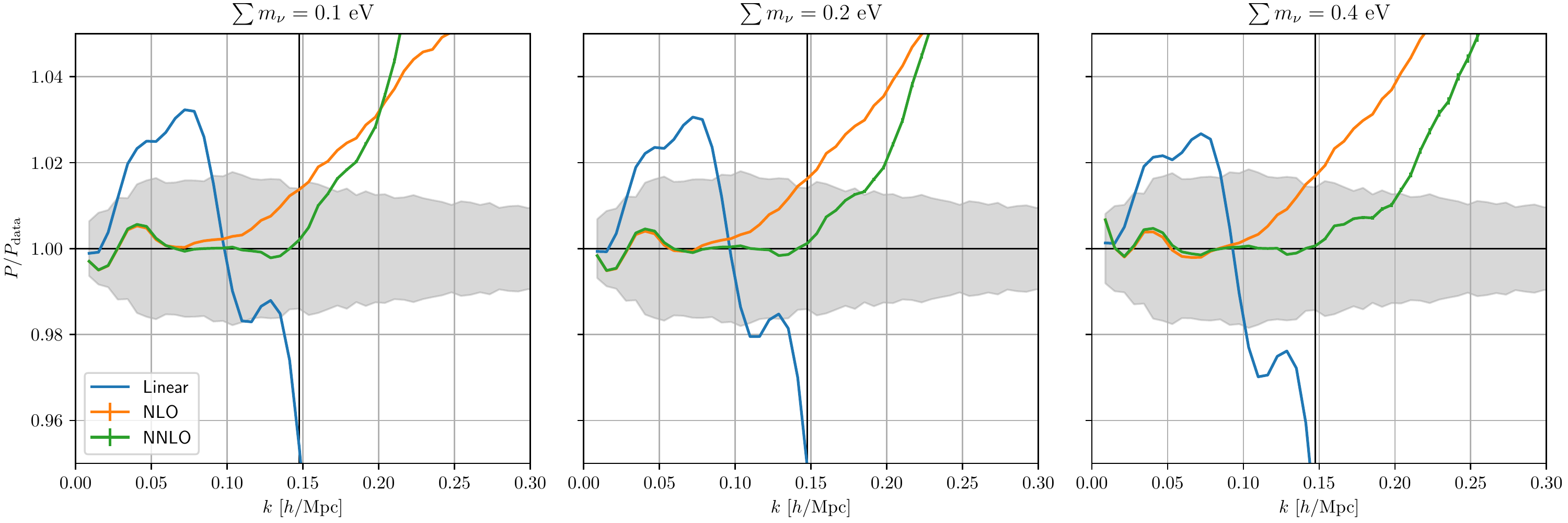}
    \caption{CDM+baryon density power spectrum predictions at linear,
    next-to-leading- and next-to-next-to-leading order in perturbation theory,
    normalized to Quijote N-body results, for three neutrino masses $\Mnu = 0.1$, $0.2$
    and $0.4~\eV$. The NLO curve correspond to the 0-parameter model and the
    NNLO curve to the 2-parameter model defined in Sec.~\ref{sec:numerics}, and
    we use a pivot scale $\kmax = 0.148\ihMpc$. The gray area indicates
    uncertainty from the N-body simulation.}
    \label{fig:pk_mass}
\end{figure}

A main result of this work is shown in Fig.~\ref{fig:pk_mass}, where we plot
the power spectrum computed in the full perturbative solution, incorporating
neutrino perturbations beyond the linear order as well as the departure from
EdS. The results are normalized to the N-body results, and we show the linear,
NLO and NNLO perturbative predictions for three neutrino masses. It is clear
that adding higher order corrections extends the wavenumber range with percent
accuracy. In particular, the NLO and NNLO results deviates by more than 1\% for
$k\gtrsim 0.12\ihMpc$ and $k\gtrsim 0.16\ihMpc$ in all cosmologies, respectively.
We refer to Sec.~\ref{sec:numerics} for further discussion, including
potential limitations related to overfitting.

Our work is structured as follows: in the next section we describe how we
capture neutrino perturbations in a two-component fluid model and compare it
to a simplified approach for the (unrenormalized) density and velocity power
spectra in Sec.~\ref{sec:bare_pk_comparison}. We detail how we renormalize the
power spectrum using an EFT approach in Sec.~\ref{sec:eft} and compare and
calibrate both models to N-body simulations in Sec.~\ref{sec:numerics}. We
conclude in Sec.~\ref{sec:conclusions}.

\section{Perturbative modelling with massive neutrinos}
\label{sec:pt_setup}

In this section we briefly review Standard Perturbation Theory (SPT) and the
generic extension of it introduced in \cite{Garny:2020ilv} that can describe
cosmologies with multiple species in addition to scale- and time-dependent
dynamics. We will apply this framework to cosmologies with massive neutrinos,
therefore we next summarize a two-component fluid setup of CDM+baryons and
neutrinos introduced in~\cite{Blas:2014} and further developed
in~\cite{Garny:2020ilv}, which can readily be captured by the extension of SPT.
Finally, we compare the two-component fluid model at the linear level to the
full neutrino Boltzmann hierarchy.

\subsection{Standard perturbation theory}

The equations of motion for the density contrast $\delta$ and velocity
divergence $\theta = \partial_i \vv^i $ (neglecting vorticity) in Fourier space
reads
\begin{align}
    \partial_{\tau} \delta(\vk,\tau) + \theta(\vk,\tau) & =
        - \int_{\vk_1,\vk_2}\delta_D (\vk - \vk_{12}) \alpha(\vk_1,\vk_2)
        \theta(\vk_1, \tau) \delta(\vk_2, \tau) \, ,
    \nonumber \\
    \partial_{\tau} \theta(\vk,\tau) + \mH \theta(\vk, \tau)
    + \frac{3}{2} \mH^2 \Omega_m \delta(\vk,\tau) & =
        - \int_{\vk_1,\vk_2}\delta_D (\vk - \vk_{12}) \beta(\vk_1,\vk_2)
        \theta(\vk_1, \tau) \theta(\vk_2, \tau) \, ,
    \label{eq:cont_euler}
\end{align}
where $\tau$ is conformal time, $\mH = \dd \ln a/\dd \tau$ the conformal
Hubble rate and $\Omega_m$ the time-dependent matter density parameter. We
introduced the shorthand notations $\vk_{12} = \vk_1 + \vk_2$ and $\int_{\vk} =
\int \dd^3 \vk$ and used $\delta_D$ to denote the Dirac delta function. The
mode coupling functions are
\begin{equation}
    \alpha(\vk_1,\vk_2) = 1 + \frac{\vk_1\cdot\vk_2}{k_1^2}\, ,\quad
    \beta(\vk_1,\vk_2) = \frac{(\vk_1 + \vk_2)^2 (\vk_1 \cdot \vk_2)}{2 k_1^2 k_2^2}\,,
\end{equation}
as usual. In SPT one assumes that the anisotropic stress of the fluid vanishes.
We also make this assumption here initially, but will relax it when discussing
an effective field theory setup in Sec.~\ref{sec:eft}.

The equations of motion can be written in a compact form after defining the
tuple $\psi = (\delta, -\theta/\mH f)$ and using $\eta = \log D$, with $D$ and
$f$ being the linear growth factor and growth rate, respectively,
thus~\cite{Bernardeau:2001qr}:
\begin{equation}
    \partial_{\eta} \psi_a(\vk,\eta) + \Omega_{ab}(\eta) \psi_b(\vk,\eta) =
    \int_{\vk_1, \vk_2} \delta_D(\vk-\vk_{12})\gamma_{abc}(\vk,\vk_1,\vk_2) \psi_b(\vk_1,\eta) \psi_c(\vk_2,\eta)\,.
    \label{eq:eom_compact_general}
\end{equation}
The matrix $\Omega_{ab}$ governing the linear evolution is given by
\begin{equation}
    \Omega_{ab}(\eta) =
    \begin{pmatrix}
        0 & -1 \\
        - \frac{3}{2} \frac{\Omega_m}{f^2} & \frac{3}{2} \frac{\Omega_m}{f^2} - 1
    \end{pmatrix}\,,
    \label{eq:omega_1F}
\end{equation}
and the only non-zero components of the non-linear vertex are
$\gamma_{121}(\vk,\vk_1,\vk_2) = \alpha(\vk_1,\vk_2)$ and
$\gamma_{222}(\vk,\vk_1,\vk_2) = \beta(\vk_1,\vk_2)$.

In SPT one typically adopts the Einstein\textendash de-Sitter (EdS)
approximation, in which $\Omega_m / f^2 = 1$ so that the $\Omega_{ab}$-matrix
becomes time-independent. This approximation greatly simplifies
Eq.~\eqref{eq:eom_compact_general}, allowing for analytic solutions order by
order. Only the decaying mode is affected by changes in the ratio
$\Omega_m/f^2$ and moreover the ratio only departs significantly from one at
late times, $z \lesssim 2$ in \LCDM\ (and also in moderate extensions).
Consequently, the EdS-approximation has been shown to work at the percent-level
for the power
spectrum~\cite{Takahashi:2008yk,Fasiello:2016qpn,Garny:2020ilv,Donath:2020abv,Fasiello:2022lff}
as well as the bispectrum~\cite{Steele:2020tak,Baldauf:2021zlt}. In particular,
in EFT analyses the departure from EdS can be largely degenerate with
counterterms, only leading to a shift in the EFT
parameters~\cite{Donath:2020abv,Baldauf:2021zlt}. In this work we consider
schemes with and without the EdS approximation, which we will specify
accordingly.

\subsection{Extension of SPT}

Following \cite{Garny:2020ilv}, we extend SPT by allowing for multiple species
in the fluid as well as allowing for a general time- and wavenumber-dependence.
More precisely, for an $N$-fluid we collect the density contrast and velocity
divergence for each component $i$ into into the field vector
$\psi_a = (\dotsc,\delta_i,\theta_i,\dotsc)$ with the index $a$ running from
$1$ to $2N$. In addition, we permit a general dependence on time and wavenumber
in the (now $2N\times 2N$) matrix describing the linear evolution
$\Omega_{ab} = \Omega_{ab}(|\vk|,\eta)$. This extension can capture multiple
models beyond \LCDM\ in addition to effective models of clustering dynamics. It
has in general no analytic solution however, hence we will mostly need to solve
the dynamics numerically.

The equations of motion \eqref{eq:eom_compact_general} can also in this case be
solved perturbatively, at each order furnished by $2N$ kernels $F_a^{(n)}$
labeled by the index $a$ at order $n$:
\begin{equation}
    \psi_a(\vk, \eta) = \sum_{n=1}^{\infty}
        \int_{\vq_1,\dotsc,\vq_n} \delta_D(\vk - \vq_{1\cdots n}) \,
        \e^{n\Delta\eta} \, F_a^{(n)} (\vq_1,\dotsc,\vq_n; \eta) \,
        \delta_0(\vq_1; \etaini)\dotsb \delta_0(\vq_n; \etaini)\, ,
        \label{eq:Fn}
\end{equation}
where $\Delta\eta\equiv \eta - \etaini$ and $\delta_0$ is an initial condition
that we discuss shortly. Note that due to the assumed non-trivial
time-dependence in the dynamics, we allow for a dependence on $\eta$ in
addition to the wavenumbers $\vq$ for the kernels. Inserting this solution into
Eq.~\eqref{eq:eom_compact_general} yields the following recursive solution at
$n$-th order in perturbation theory:
\begin{align}
    \lefteqn{
        (\partial_{\eta} + n) F_a^{(n)}(\vq_1,\dotsc,\vq_n;\eta) + \Omega_{ab}(k,\eta)
    F_b^{(n)}(\vq_1,\dotsc,\vq_n;\eta) }\nn\\
    & = &
    \sum_{m=1}^{n-1} \Big[\gamma_{abc}(\vk,\vq_{1\dotsb m},\vq_{m+1\dotsb n})
    F_b^{(m)}(\vq_1,\dotsc,\vq_m;\eta)
    F_c^{(n-m)}(\vq_{m+1},\dotsc,\vq_n;\eta)\Big]_{\rm sym.}
    \, .\quad
    \label{eq:eom_kernels}
\end{align}
Here, $\vk = \sum_i \vq_i$, and the right hand side is understood to be
symmetrized with respect to all permutations exchanging momenta in the
$\{\vq_1,\dotsc,\vq_m\}$ set with momenta in the $\{\vq_{m+1},\dotsc,\vq_n\}$
set and normalized to the number of permutations.

Note that setting $N=1$ and using the $\Omega_{ab}$-matrix from
Eq.~\eqref{eq:omega_1F} with $\Omega_m/f^2 = 1$ (EdS approximation) in
Eq.~\eqref{eq:eom_kernels}, we recover in the limit $\etaini \to - \infty$ the
usual kernel recursion relations with the replacements
$F_1^{(n)}\to F_n$ and $F_2^{(n)}\to G_n$.

We still need to specify suitable initial conditions in order to solve the
above equations. Taking $\etaini$ after recombination but long before
non-linearities become important at the scales of interest, we assume that the
initial conditions for each fluid component is correlated, so that
\begin{equation}
    \psi_a(\vk, \etaini) = F_a^{(1)}(k, \etaini) \, \delta_0(\vk)\,,
\end{equation}
which holds for adiabatic initial conditions. Furthermore, we assume that
$\delta_0$ is a Gaussian random field, so that we only need to specify the
initial linear power spectrum
$\langle\delta_0(\vk)\delta_0(\vk')\rangle=\delta_D(\vk+\vk')P_0(k)$ in order
to compute correlations of $\psi_a$'s. The initial linear kernels $F_a^{(1)}(k,
\etaini)$ impose the relative normalization for each perturbation and
wavenumber. Deep in the linear regime $\etaini\to\infty$ the higher order
initial kernels can be set to zero. In practice however, using $\etaini$ after
recombination, we find that those $n>1$ initial conditions work poorly because they
excite transient solutions that do not entirely decay by $\eta = 0$ ($z = 0$).
We return to this issue below.

We are ultimately interested in the statistical properties of the fields
$\psi_a(\vk,\eta)$, in particular auto- and cross power spectra
\begin{equation}
    \langle\psi_a(\vk,\eta)\psi_b(\vk',\eta)\rangle
        =\delta_D(\vk+\vk')P_{ab}(k,\eta)\,.
    \label{eq:Pab}
\end{equation}
The perturbative expansion \eqref{eq:Fn} combined with the Wick theorem yields
as a result the loop expansion of the power spectrum
\begin{equation}
    P_{ab}(k,\eta)=P_{ab}^{\rm lin}(k,\eta)+P_{ab}^{\oneloop}(k,\eta)+P_{ab}^{\twoloop}(k,\eta)+\dots\,.
    \label{eq:ps_loop}
\end{equation}
To compute loop corrections, we employ the numerical algorithm described
in~\cite{Garny:2020ilv} (see also~\cite{Blas:2013bpa, Blas:2013aba,
Blas:2015tla}). In short, at $L$-loop it consists of integrating over $L$ loop
momenta using Monte Carlo integration (with CUBA~\cite{Hahn:2004fe}) where at
every integration point a set of $(2L+1)$-order kernels needs to be evaluated
using the recursion relation \eqref{eq:eom_kernels}. In general, there is no
analytic solution of Eq.~\eqref{eq:eom_kernels}, therefore we solve for the
kernels numerically. We refer to~\cite{Garny:2020ilv} for further details on
the algorithm.

\subsection{Two-component fluid: CDM+baryons and massive neutrinos}

We will employ the hybrid two-component fluid setup described in
\cite{Garny:2020ilv} to model structure formation in massive neutrino
cosmologies. In the following, we repeat the main elements of this setup for
convenience, but refer to \cite{Garny:2020ilv} for details of the
implementation.

In the hybrid two-component fluid model, the system is described linearly by
the full Boltzmann hierarchy until some intermediate redshift $\zmatch$, after
which the evolution is mapped onto a two-component fluid, suitable for
computing non-linear corrections (see also \cite{Blas:2014}). Baryons and CDM
comprise jointly one fluid component, which is coupled to the second component,
the neutrinos, via gravity. A fluid description of massive neutrinos is
suitable at late times because the coupling to higher moments in the Boltzmann
hierarchy is suppressed by powers of $T_{\nu}/m_{\nu}$. We may therefore follow
only the lowest moments of the hierarchy for the neutrinos, with an effective
sound velocity to capture free-streaming. We show below that this only leads to
sub-percent differences compared to the full hierarchy. The CDM+baryon
component is treated both as a perfect, pressureless fluid (SPT) as well as an
imperfect fluid with corrections to the effective stress tensor (EFT).

We consider three massive neutrino cosmologies with $M_{\nu} \equiv \sum_i
m_{\nu_i} = 0.1$, $0.2$ and $0.4~\eV$. In all models we assume that the
neutrinos are degenerate, each with mass $M_{\nu}/3$. This approximation works
well because cosmological observables are very insensitive to the neutrino mass
hierarchy except for the total mass, see e.g.~\cite{Archidiacono:2020dvx}.
Furthermore, we choose a matching redshift $\zmatch = 25$, which for the
neutrino masses of consideration is well after the non-relativistic transition
\begin{equation}
    z_{\mathrm{nr}} = 1890 \frac{m_{\nu}}{1~\mathrm{eV}}\,,
    \label{eq:z_nr}
\end{equation}
in addition to being sufficiently earlier than the point at which
non-linearities become important, $z \lesssim 10$.

The weighted density contrast for the CDM+baryon fluid component reads
\begin{equation}
    \delta_{\cb} = \frac{ f_{\mathrm{b}}\delta_{\mathrm{b}} + f_{\cdm} \delta_{\cdm}}
        {f_{\mathrm{b}} + f_{\cdm}}\,,
    \label{eq:delta_cb}
\end{equation}
with $f_i \equiv \Omega_i/\Omega_{\mathrm{tot}}$, and similarly for the velocity
divergence. We collect the density contrasts and (suitably rescaled) velocity
divergences of both fluid components into a vector~%
\footnote{Due to neutrino free-streaming the linear growth functions $D(z)$ and
$f(z)$ become scale dependent in massive neutrino cosmologies. To avoid a
time-parameterization and rescaling in Eq.~\eqref{eq:2F_psi} dependent on
scale, we redefine $\eta = \ln \left(D \big|_{\fnu = 0}\right)$ and
$f = f \big|_{\fnu = 0}$ in the two-fluid setup.%
}
\begin{equation}
    \psi_a = \left( \delta_{\cb}\,,\quad - \frac{\theta_{\cb}}{\mH f}\,,\quad
    \delta_{\nu}\,,\quad - \frac{\theta_{\nu}}{\mH f} \right).
    \label{eq:2F_psi}
\end{equation}
The equations of motion for the two-component fluid are given by
Eq.~\eqref{eq:eom_compact_general} with
\begin{equation}
    \Omega_{ab}(k, \eta) =
    \begin{pmatrix}
        0 & -1 & 0 & 0 \\
        - \frac{3}{2} \frac{\Omega_m}{f^2} (1 - \fnu) &
        \frac{3}{2} \frac{\Omega_m}{f^2} - 1 &
        - \frac{3}{2} \frac{\Omega_m}{f^2} \fnu & 0 \\
        0 & 0 & 0 & -1 \\
        - \frac{3}{2} \frac{\Omega_m}{f^2} (1 - \fnu) &
        0 &
        - \frac{3}{2} \frac{\Omega_m}{f^2} [\fnu - k^2 \cseff^2(k,\eta) ] &
        \frac{3}{2} \frac{\Omega_m}{f^2} - 1
    \end{pmatrix}\,,
    \label{eq:2F_omega}
\end{equation}
and the non-zero components of the vertex $\gamma_{abc}$ being
\begin{align}
    \gamma_{121}(\vk,\vk_1,\vk_2) =
    \gamma_{343}(\vk,\vk_1,\vk_2) \equiv \alpha(\vk_1,\vk_2)\,, \nonumber \\
    \gamma_{222}(\vk,\vk_1,\vk_2) =
    \gamma_{444}(\vk,\vk_1,\vk_2) \equiv \beta(\vk_1,\vk_2)\,.
    \label{eq:2F_gamma}
\end{align}

In the part of the evolution matrix $\Omega_{ab}$ describing the neutrino dynamics, we
introduced the effective neutrino sound velocity $c_{s,\mathrm{eff}}^2$. This
term captures the free-streaming of the neutrinos $c_s^2 = \delta
P_{\nu}/\delta \rho_{\nu}$ as well as the neutrino anisotropic stress
$\sigma_{\nu}$:
\begin{equation}
    c_{s,\mathrm{eff}}^2(k, \eta)
    \equiv \frac{1}{\kFS^2(k, \eta)}
    = \frac{2}{3 \Omega_m \mH^2}
    \left[
        c_s^2(k,\eta) - \frac{\sigma_{\nu}(k,\eta)}{\delta_{\nu}(k,\eta)}
    \right] \, ,
    \label{eq:kFS}
\end{equation}
where we defined the associated effective free-streaming wavenumber $\kFS$. We compute
the terms in the bracket in linear theory, however in a manner that is informed
about the complete neutrino distribution function, including all higher moments
(see \cite{Garny:2020ilv}).

The equations for the two-component fluid are precisely of the form that can be
solved by the extension of SPT discussed above. We expand $\psi_a$ in powers of
$\delta_0 = \delta_{\cb}(\etamatch)$ using Eq.~\eqref{eq:Fn}. The accompanying
kernels are solutions of Eq.~\eqref{eq:eom_kernels}, with $\Omega_{ab}$ and
$\gamma_{abc}$ from Eqs.~\eqref{eq:2F_omega} and \eqref{eq:2F_gamma}.
We find that we need to be cautious in choosing initial conditions for the
kernel hierarchy at the matching redshift; one easily excites transient
solutions that do not entirely decay away by $\eta = 0$. Ultimately, we opt for
numerically finding the growing mode solution for the kernels with fixed
$\Omega_{ab}(k) = \Omega_{ab}(k,\etamatch)$ and using this as initial
condition. This method is described more comprehensively
in~\cite{Garny:2020ilv}. We check that this initial condition matches at the
linear level the transfer functions from the Boltzmann solver
CLASS~\cite{Lesgourgues:2011re} for the wavenumbers of interest, and that we
obtain the same linear power spectrum at $z = 0$ (see below).

The total matter power spectrum is the sum of the weighted auto spectra for
each component as well as the cross-spectrum,
\begin{equation}
    P_{\mm} = (1 - \fnu)^2 P_{\cb,\cb} + 2 (1 - \fnu) \fnu P_{\cb,\nu}
        + \fnu^2 P_{\nu,\nu}\,.
    \label{eq:PS_mm}
\end{equation}
The neutrino energy fraction is $\fnu = 0.75$, $0.15$ and $0.3\%$ (constant for
$z \ll z_{\mathrm{nr}}$) for the three models with $\Mnu = 0.1$, $0.2$ and
$0.4~\eV$, respectively. For wavenumbers larger than the neutrino
free-streaming scale $\kFS \sim 0.01~h/\Mpc$, neutrinos are hardly captured in
gravitational wells and the neutrino spectra are suppressed compared to
$P_{\cb,\cb}$. In total, the first term in Eq.~\eqref{eq:PS_mm} gives the
dominant contribution to the matter power spectrum. Nevertheless, via the
backreaction on the gravitational potential, the presence of massive neutrinos
leads to a reduction of growth of the CDM+baryon fluid. At the linear level,
the relative suppression of the matter power spectrum between a model with
massive neutrinos and one without (with $\Omega_{\cdm}$ tuned so that
$\Omega_m$ is the same) is the well-known $-8\fnu$~\cite{Lesgourgues:2006nd}.

\begin{figure}[t]
    \centering
    \includegraphics[width=0.6\textwidth]{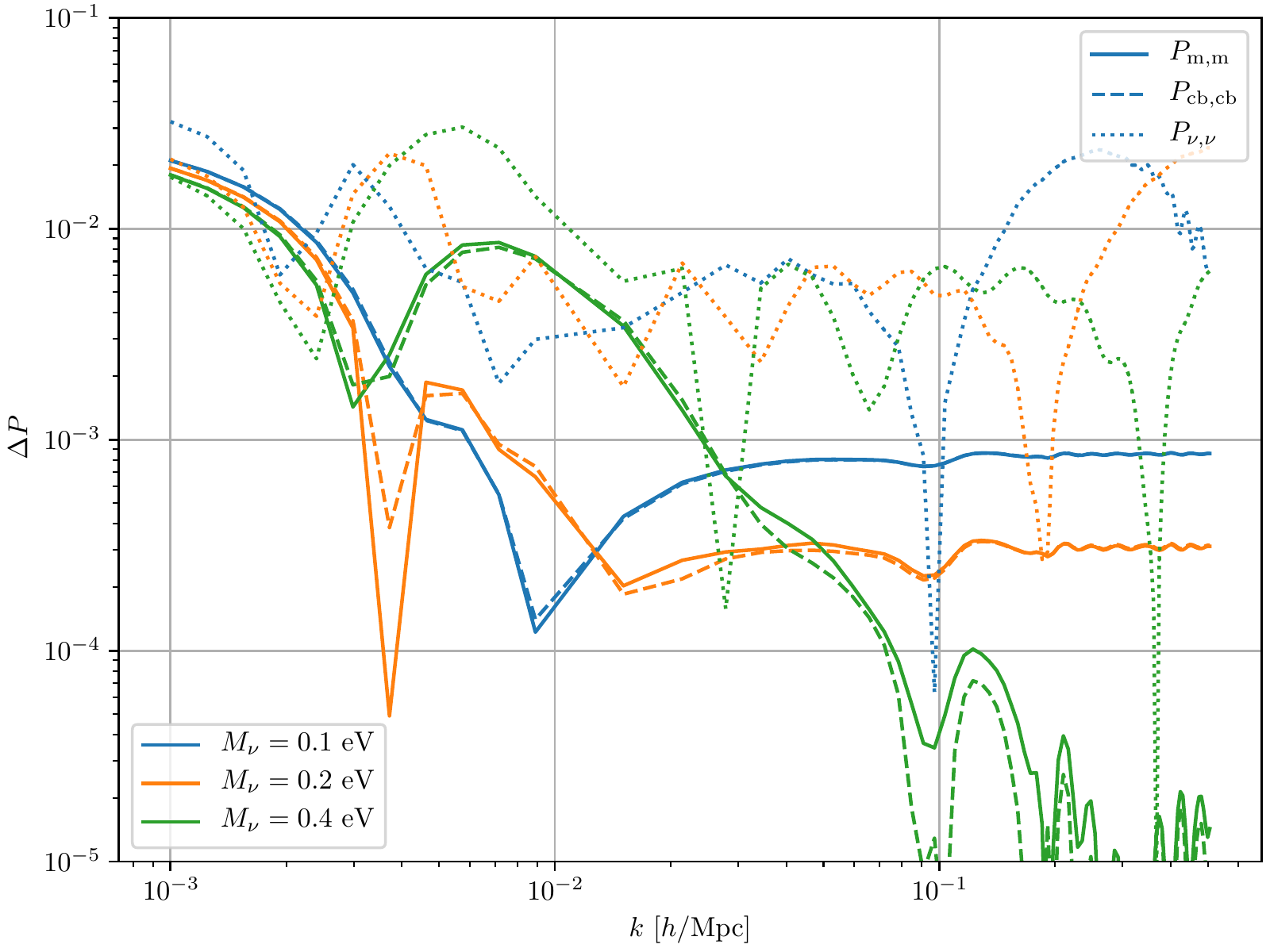}
    \caption{Relative difference between the linear power spectra computed
    using the two-fluid model and using the full Boltzmann hierarchy for the
    neutrino perturbation (CLASS), $\Delta P = | 1 - P/P_{\mathrm{CLASS}} | $.
    The solid, dashed and dotted lines correspond to $P_{\mm}$, $P_{\cb,\cb}$
    and $P_{\nu,\nu}$, respectively, while the blue, orange and green lines
    correspond to $\Mnu = 0.1$, $0.2$, and $0.4~\eV$, respectively.}
    \label{fig:lin_comparison}
\end{figure}

In Fig.~\ref{fig:lin_comparison}, we compare the two-fluid scheme with the full
solution using the Boltzmann hierarchy (CLASS, with maximum neutrino multipole
$l_{\mathrm{max}} = 17$) at the linear level. The matter power spectra have
sub-permille agreement for $k > 0.03~h/\Mpc$ for all neutrino masses. For very
low wavenumber, $k\to10^{-3}~h/\Mpc$ there are percent-level deviations. These
arise because at the matching redshift $\zmatch = 25$, the wavenumbers have not
yet completely reached the growing mode after entering the horizon, while the
two-fluid model initial condition assumes that all scales already reside in the
growing mode. In addition, for scales close to Hubble, relativistic corrections
become important. Nevertheless, these large scales have negligible impact on
loop corrections for a realistic power spectrum.

The neutrino auto power spectrum agrees better than 1\% for scales $k \sim
0.01$ \textendash\ $0.1~h/\Mpc$. On smaller scales the difference increase,
however in this region errors from the truncation of the Boltzmann hierarchy
(choice of $l_{\mathrm{max}}$) also become significant. We see that the
deviation of the neutrino spectrum has a minimal effect on the total matter
power spectrum, whose dominant contribution is the CDM+baryon component.

\section{Effect of neutrino perturbations on density and velocity spectra}
\label{sec:bare_pk_comparison}

Having set up the formalism and procedure for computing non-linear corrections
taking neutrino perturbations into account beyond the linear level, we proceed in
this section to comparing it to simplified treatments. We consider the bare
density- and velocity power spectra, and leave a comparison of the renormalized
power spectra to Sec.~\ref{sec:numerics}. Furthermore, we assess to which
extent the effect of free-streaming neutrinos is degenerate with the primordial
amplitude of fluctuations at one- and two-loop on BAO scales.

The schemes for describing the growth of structure in presence of massive
neutrinos that we will consider in this work in increasing order of complexity are:
\begin{enumerate}
    \item \textbf{EdS-SPT scheme (1F).} The impact of neutrinos is only taken into
        account at the linear level. Therefore, the only species modelled
        non-linearly by the fluid Eqs.~\eqref{eq:eom_compact_general} is
        CDM+baryons (joint fluid). In addition, the ratio $\Omega_m/f^2$ is
        approximated to 1 so that analytical EdS-SPT kernels may be used. This
        approach is frequently used in the literature because of its
        simplicity and efficiency, the latter being particularly favorable for
        data analysis and parameter scans.
    \item \textbf{One-fluid scheme (1F+).} Equivalent to the 1F-scheme, but
        relaxing the EdS approximation by including the time-dependence of
        $\Omega_m/f^2$. This scheme is included in order to estimate to what
        degree the EdS approximation is the source of inaccuracies when
        comparing 1F and 2F.
    \item \textbf{Two-fluid scheme (2F).} Baryons+CDM and neutrinos are both
        modelled beyond the linear theory in the two-component fluid setup. The
        time- and scale-dependence of the dynamics both arising from the
        departure from EdS at late times as well as neutrino free-streaming are
        captured by solving the kernel hierarchy~\eqref{eq:eom_kernels}
        numerically, with the linear evolution and non-linear vertices given by
        Eqs.~\eqref{eq:2F_omega} and \eqref{eq:2F_gamma}, respectively.
\end{enumerate}
For the 1F and 1F+ schemes, we use the linear power spectrum at $z = 0$ (from
CLASS) as input to the loop correction calculations. The 2F scheme however,
captures also accurately the linear evolution (see discussion above), thus we
take the linear power spectrum at $\zmatch = 25$ as input. The increasing
complexity of the different schemes leads to longer processing times; in
Table~\ref{tab:cpu_times} we list the approximate times to compute
one integration point on one core on a laptop at one- and two-loop in each
scheme. A typical calculation of a loop correction for a single external
wavenumber entails $\mathcal{O}(10^6)$ integral evaluations.

\begin{table*}[t]
    \centering
    \caption{Approximate time in microseconds to compute one integration point
    (the integrand for a fixed set of integration variables) at one- and
    two-loop in the different schemes. Measured on a laptop using one core.}
    \label{tab:cpu_times}
    {\def\arraystretch{1.55}
    \begin{tabular}{l|rrr}
        \hline
                   & EdS-SPT (1F) & One-fluid (1F+) & Two-fluid (2F) \\
        \hline
        1-loop     & 3            & 620             & 1600           \\
        2-loop     & 50           & 5400            & 15000          \\
    \end{tabular}
    }
\end{table*}

\subsection{Comparison}

To address the accuracy of treating neutrinos only linearly as well as the EdS
approximation, we compare the 1F and 2F schemes for the CDM+baryon density and
velocity power spectra at $z = 0$. We consider the power spectrum both at NLO
(linear + one-loop) and NNLO (linear + one-loop + two-loop), without EFT
corrections (we repeat the comparison including these in
Sec.~\ref{sec:numerics}).

\begin{figure}[t]
    \centering
    \includegraphics[width=\linewidth]{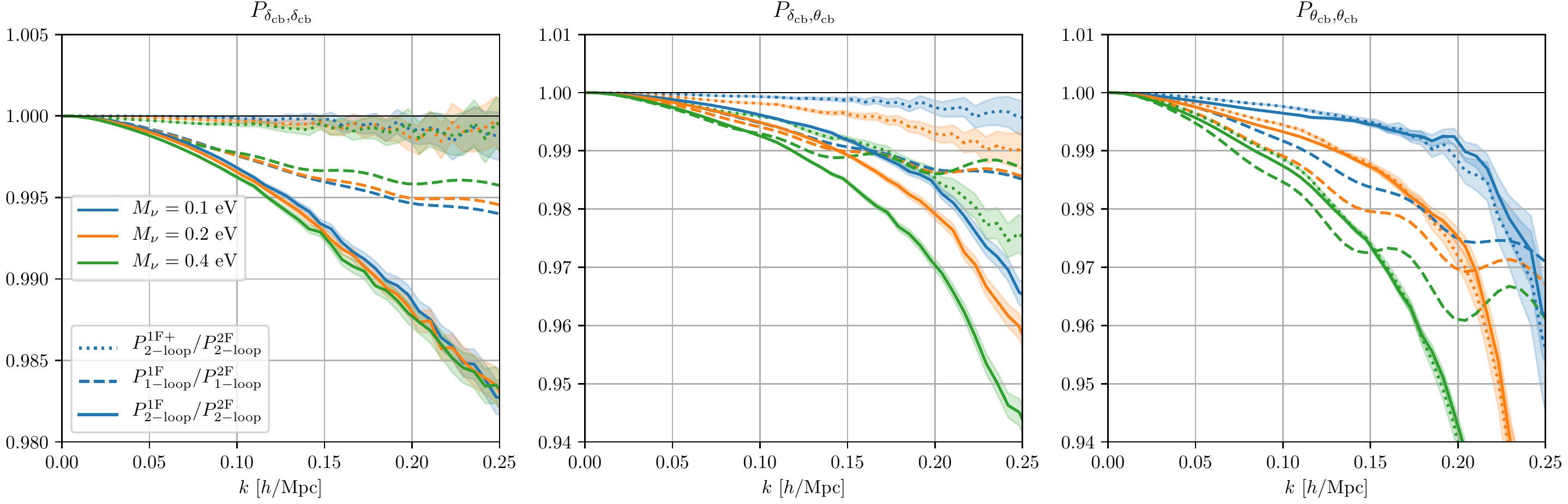}
    \caption{Comparison of the 1F, 1F+ and 2F schemes for the CDM+baryon
    density and velocity spectra. We show the power spectrum at one-loop
    (dashed lines) and two-loop (solid lines) in the 1F scheme, divided by that
    from the 2F scheme. The dotted lines correspond to the 1F+ scheme
    normalized to the 2F scheme, at two-loop order. Shaded regions indicate
    uncertainty from the Monte Carlo integration. \emph{Left:} density auto
    spectrum, \emph{middle:} density-velocity cross spectrum and \emph{right:}
    velocity auto spectrum.
    }
    \label{fig:loop_corr}
\end{figure}

The comparison is presented in Fig.~\ref{fig:loop_corr}. In the leftmost panel
we show the density auto spectrum $P_{\delta_{\cb},\delta_{\cb}}$ in the 1F and
1F+ schemes, normalized to the 2F result. At one-loop (dashed lines), the 1F
scheme deviates at most by $0.5\%$ from the full 2F solution, while at two-loop
(solid lines) the deviation exceeds one percent at $k\simeq 0.17~\ihMpc$ and
grows as the two-loop correction becomes increasingly significant. A similar
comparison was done for different neutrino masses in \cite{Garny:2020ilv}, and
it is reassuring to see consistent results. Since the deviations are roughly
independent of neutrino mass and there is insignificant difference between the
1F+ (EdS approximation relaxed) and 2F schemes at two-loop (dotted lines), we
conclude that the main source of inaccuracy for the density spectrum comes from
the EdS approximation.%
\footnote{Our results for the difference between EdS (1F) and exact
time-dependent kernels (1F+) at one- and two-loop are consistent with those
of~\cite{Fasiello:2022lff}.}

The middle and right panels display the same comparison for the cross spectrum
$P_{\delta_{\cb},\theta_{\cb}}$ and velocity spectrum
$P_{\theta_{\cb},\theta_{\cb}}$. In contrast to the density spectrum, they show
a clear dependence on the neutrino mass in the fractional difference. Moreover,
the deviations of the 1F and 1F+ schemes from the full 2F solution
are very similar for $P_{\theta_{\cb},\theta_{\cb}}$, indicating that the main
error comes from neglecting the effect of non-linear neutrino perturbations. It
is not too surprising that this effect is larger on the fluid velocity than the
density contrast: for scales smaller than the free-streaming scale the growing
mode for the CDM+baryon component is approximately $(1, 1 - 3/5\, \fnu)$ in the
2F scheme, while it remains $(1,1)$ in 1F (and 1F+). Therefore, there is an
extra suppression for the velocity field compared to the density field, in
addition to the change of the growth factor. Nevertheless, the deviation for the
smallest neutrino mass $\Mnu = 0.1~\eV$ is less than a percent at two-loop for
$k \lesssim 0.15$ and $0.2~\ihMpc$ for the cross- and velocity spectra,
respectively. We note that for wavenumbers $k \gtrsim 0.2~\ihMpc$ the two-loop
correction is of the order of the one-loop correction, signalling the breakdown
of perturbation theory. Finally, for the velocity spectra it is curious to see
that the deviations of 1F to 2F for the one- and two-loop corrections go in
opposite directions, thus adding the two-loop seemingly improves the agreement.

\subsection{Degeneracy with amplitude of fluctuations}

The suppression of power on scales smaller than the neutrino free-streaming
scale is largely degenerate with a change of the clustering amplitude $A_s$.
This may pose a major limitation on constraining the neutrino mass with
large-scale structure%
~\cite{Viel:2010bn,Marulli:2011he,Villaescusa-Navarro:2013pva,Archidiacono:2016lnv}.
The complication is somewhat mitigated by measuring at different redshifts,
including higher-order statistics~\cite{Hahn:2019zob,Hahn:2020lou} and by
redshift-space distortions~\cite{Villaescusa-Navarro:2017mfx}. Moreover, the
degeneracy can be broken by including other probes, such as CMB data.

\begin{figure}[t]
    \centering
    \includegraphics[width=\linewidth]{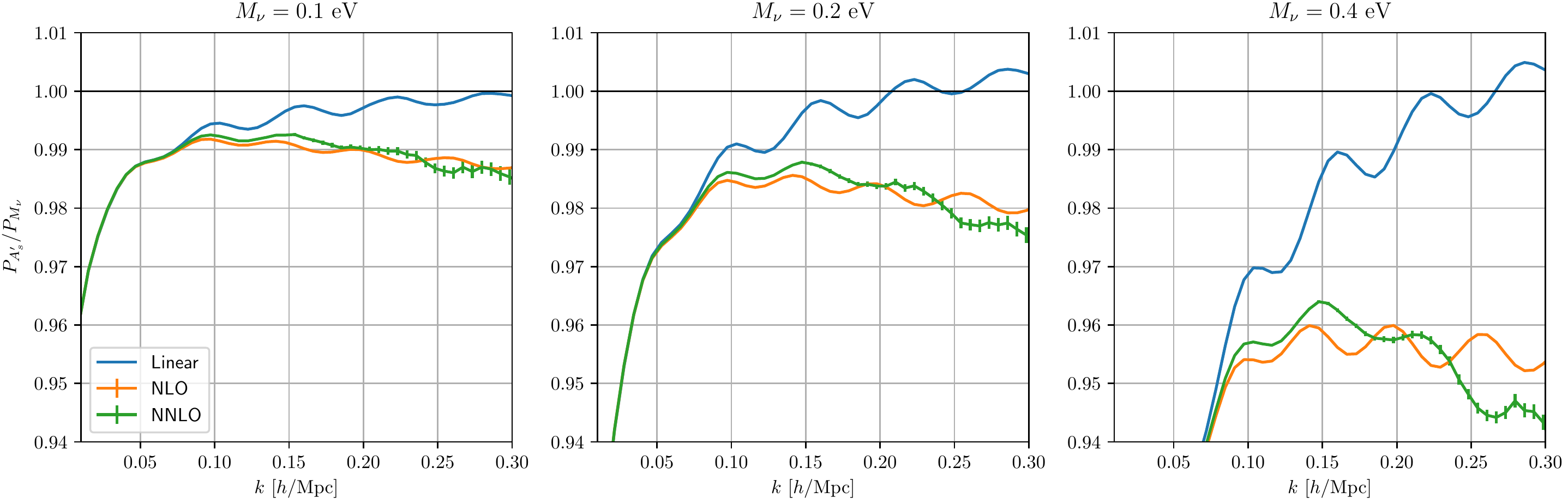}
    \caption{Comparison of the total matter power spectrum in massive neutrino
    models and massless models with the amplitude of fluctuations changed. We
    show massless models with $A_s' = (1 - 8\fnu)A_s$ normalized to massive
    models with amplitude $A_s$ for the linear (blue), NLO (orange) and NNLO
    (green) power spectrum. To isolate the effect of the neutrinos, we compute
    the massless model in the 1F+ scheme and the massive model in the 2F
    scheme. Errorbars indicate uncertainty from Monte Carlo integration.}
    \label{fig:As_deg}
\end{figure}

In this light, we check to what extent we can reproduce our results for the 2F
matter power spectra in massive neutrino models using massless models by
adjusting $A_s$. To isolate the effect of neutrino suppression and overall
amplitude, we use the 1F+ scheme for the massless models. Furthermore, we
remove completely the double-hard limit of the two-loop corrections, which
otherwise yields a large contribution at the scales of interest (after
renormalization this contribution is considerably reduced). In
Sec.~\ref{sec:eft} we go into details on this procedure and include EFT
corrections to renormalize the power spectrum. For simplicity, we set $A_s' =
(1 - 8\fnu) A_s$ to mimic neutrino free-streaming suppression in the massless
models. The results are shown in Fig.~\ref{fig:As_deg}. For the lowest neutrino
mass, the massive and modified massless models agree within a percent. As the
characteristic free-streaming wavenumber increases for larger neutrino mass, it
becomes more difficult to reconstruct the shape of the massive neutrino
non-linear power spectrum with a change of $A_s$. The shape dependence of the
linear suppression is more prominent at $k\simeq 0.1$ -- $0.2\ihMpc$ and due to
mode-coupling, the loop-corrections are affected beyond just a rescaling by the
presence of the neutrinos.

\section{Effective field theory setup}
\label{sec:eft}

In order to precisely assess the performance of the simplified treatment of
neutrinos (1F) and the full two-fluid solution (2F) versus N-body simulations,
we need to take into account EFT corrections. In this section we elaborate on
the EFT setup we employ for obtaining a renormalized CDM+baryon power spectrum
at next-to-next-to-leading order.

In the effective theory approach, long wavelength
fluctuations are modelled perturbatively while systematically taking into
account the impact of short wavelengths~\cite{Baumann:2010tm,Carrasco:2012cv}.
This is done by coarse-graining the fluid, leading to a modified Euler equation
with an effective stress tensor that in addition to the microscopic
stress includes corrections from coarse-grained products of short fluctuations.
In the notation introduced in Sec.~\ref{sec:pt_setup}, the equations of
motion~\eqref{eq:eom_compact_general} in the two-fluid model are modified as
\begin{align}
    \partial_{\eta} \psi_a(\vk,\eta) + \Omega_{ab}(\vk,\eta) \psi_b(\vk,\eta) & =
    \int_{\vk_1, \vk_2} \delta_D(\vk-\vk_{12})
        \gamma_{abc}(\vk,\vk_1,\vk_2) \psi_b(\vk_1,\eta) \psi_c(\vk_2,\eta)
    \nonumber \\
    & \phantom{= }
    \label{eq:eft_eom}
    + \delta_{a2}^{(K)}\, \tau_{\theta_{\cb}}(\vk)
    + \delta_{a4}^{(K)}\, \tau_{\theta_{\nu}}(\vk)\,,
\end{align}
where $\delta_{ij}^{(K)}$ is the Kronecker delta, and the effective stress terms
are defined equivalently for the CDM+baryon and neutrino fluid in real space as
\begin{equation}
    \tau_{\theta} = \partial_i \frac{1}{1 + \delta} \partial_j \tau^{ij}\,.
\end{equation}
The effective stress tensor $\tau^{ij}$ can in the effective theory be split
into a deterministic and stochastic contribution; the deterministic part can be
modelled in terms of the perturbative solution, while the stochastic part is
uncorrelated with the long modes and must be modelled statistically. Due to
momentum conservation, the deterministic part must scale as the density
contrast times $k^2$ as $k\to 0$, while the stochastic power spectrum vanishes
faster than $k^4$ in this limit~\cite{Mercolli:2013bsa,Abolhasani:2015mra}.
Therefore, we neglect the stochastic contribution in this work.

In the effective theory the deterministic part of the stress tensor is not
described from first principles, but written down as a sum of all possible
operators allowed by symmetries, with a priori unknown EFT coefficients that
must be fitted to simulations or marginalized over in data analysis. Due to
Galilean invariance and the equivalence principle, the operators are restricted
to gradients and products of the \emph{building blocks}
$\partial^i\partial^j\Phi$ and $\partial^i\vv^j$, where $\Phi$ is the rescaled
gravitational potential satisfying $\Delta\Phi = \delta$. At leading order in
gradients and number of factors of fields, the effective stress term
is~\cite{Abolhasani:2015mra}
\begin{equation}
    \tau_{\theta} \big|_{\mathrm{1}} =
    d_{\delta}^2\, \Delta \delta^{(1)}
    +
    d_{\theta}^2\, \Delta \theta^{(1)}\,,
    \label{eq:tau_lo}
\end{equation}
where $d_{\delta}^2$ and $d_{\theta}^2$ are (time-dependent) EFT coefficients,
and $\delta^{(1)}$ and $\theta^{(1)}$ are the linear solutions in perturbation
theory. In massless neutrino cosmologies, they are related by
$\delta^{(1)} = - \theta^{(1)}/\mH f$, so that the sum above can be reduced to
$d^2\Delta\delta_{(1)}$ with $d^2 = d_{\delta}^2 - d_{\theta}^2/\mH f$. This is
not the case for massive neutrino models, due to free-streaming, therefore we
in principle need to write down
\begin{equation}
    \tau_{\theta_{\cb}} \big|_{\mathrm{1}} =
    \sum_{i=1}^{4}
    d_{\cb,i}^2 \, \Delta \psi_i^{(1)}
    \qquad\text{and}\qquad
    \tau_{\theta_{\nu}} \big|_{\mathrm{1}} =
    \sum_{i=1}^{4}
    d_{\nu,i}^2 \, \Delta \psi_i^{(1)}
\end{equation}
in the two-fluid model with EFT coefficients $d_{\cb,i}$, $d_{\nu,i}$. This leads to a
counterterm CDM+baryon density field
\begin{equation}
    \tilde{\psi}_1^{(1)} \equiv \tilde{\delta}_{\cb}^{(1)} =
    k^2 \sum_{i=1}^{4}
    \mathsf{d}_{i}^2 \, \psi_i^{(1)}
    \label{eq:delta_counter}
\end{equation}
in Fourier space. Each coefficient $\mathsf{d}_i$ depends on a linear
combination of \emph{both} sets of coefficients $d_{\cb,i}$, $d_{\nu,i}$
because the linear propagation after the insertion of the sources
$\tau_{\theta_{\cb}}$ and $\tau_{\theta_{\nu}}$ mixes the contributions from
the four components.

\begin{figure}[t]
   \centering
        \includegraphics[width=0.8\textwidth]{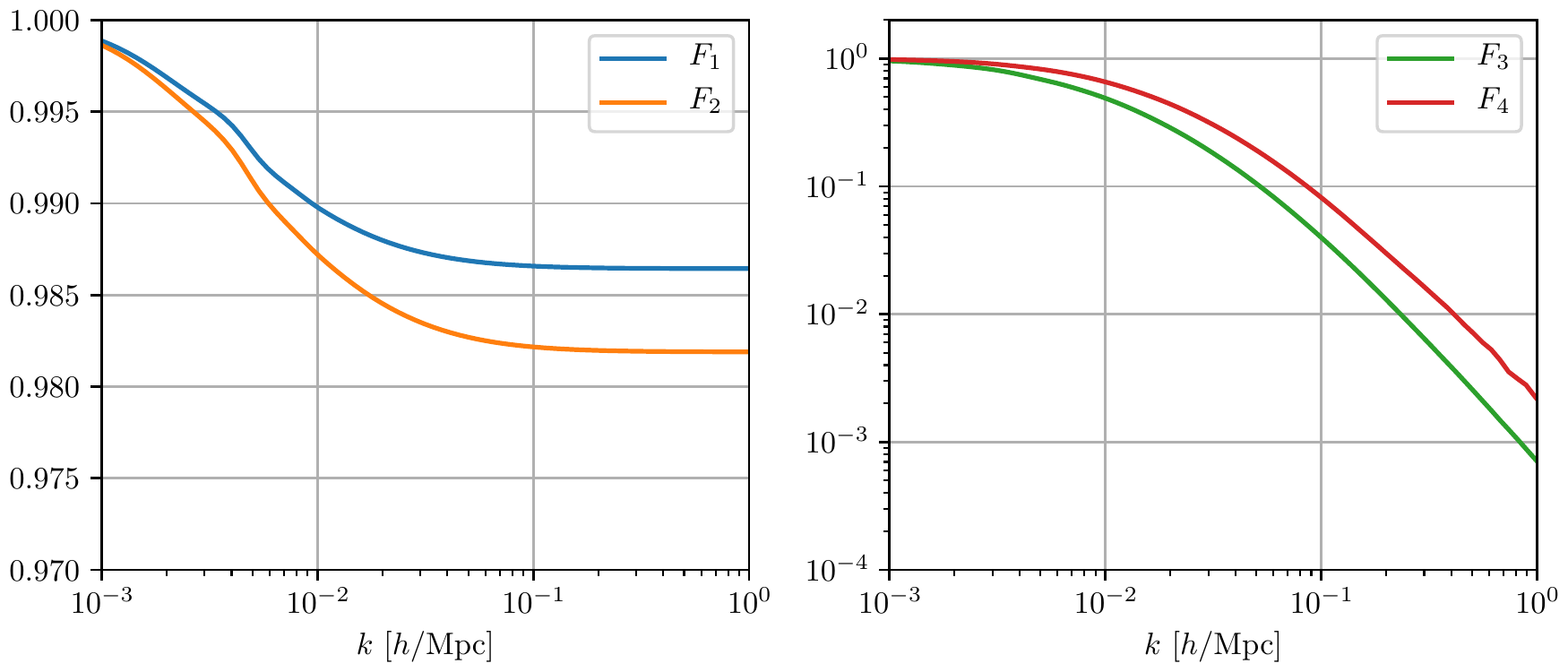}
    \caption{Linear kernels $F_a^{(1)}$ at $z = 0$ for $\Mnu = 0.1~\eV$.}
    \label{fig:F1}
\end{figure}

In practice, we are interested in weakly non-linear scales $k \simeq
0.1\ihMpc$, and assume a large scale separation $\kFS \ll k \ll \kNL$, where
the non-linear scale is $\kNL \sim 0.3\ihMpc$. As we return to below, this
assumption is justified for $\Mnu = 0.1$ and $0.2~\eV$ but begins to break down
for $\Mnu = 0.4~\eV$. In particular, the linear CDM+baryon density contrast
and (rescaled) velocity divergence can be related by a constant factor far
below the free-streaming scale: we have approximately
$\psi_2^{(1)} = (1 - 3/5\,\fnu) \psi_1^{(1)}$, see the left panel of
Fig.~\ref{fig:F1} (the linear solutions are $\psi_i^{(1)} \propto F_i^{(1)}
\delta_{\cb}(\etaini)$). Furthermore, based on the expectation that
$\mathsf{d}_i^2/\kNL^2 = \mathcal{O}(1)$ and that the neutrino perturbations
are more than an order of magnitude smaller than the CDM+baryon ones around the
scales of interest -- see the right panel of Fig.~\ref{fig:F1} -- we neglect
the contribution from the $\psi_3^{(1)}$- and $\psi_4^{(1)}$-terms. Note also
that the sound velocity in the microscopic part of the $\tau_{\nu}^{ij}$ stress
tensor is already captured in the 2F model, as defined in
Eq.~\eqref{eq:kFS}. In total, the leading CDM+baryon counterterm density
can be written as
\begin{equation}
    \tilde{\psi}_1^{(1)}(\eta) = \e^{2\Delta\eta}\gamma_1(\eta) \, k^2 \psi_1^{(1)}(\etaini)\,,
    \label{eq:delta_counter_fin}
\end{equation}
where $\gamma_1$ is an EFT parameter and we extracted the overall growth
$\e^{2\Delta\eta}$ for convenience. Since we will determine $\gamma_1$ via
calibration to N-body simulations, its explicit relation to the coefficients
$\mathsf{d}_i$ and kernels $F_i^{(1)}(\eta)$ is not important.

\subsection{Renormalization of the one-loop correction}
\label{subsec:L1_ren}

To assess whether the counterterms from the effective stress term in the EFT
can cure the cutoff-dependence of SPT (also in the 2F scheme), we study its UV
sensitivity, starting with the one-loop correction. In the EdS approximation
(1F scheme), the leading contribution (i.e.\ at leading power in the gradient
expansion $k^2/\kNL^2$) to the SPT one-loop power spectrum for hard loop
momentum $q\to\infty$ is
\begin{equation}
    P_{1L}^{h}(k,\eta) =
    - 2 \e^{4\Delta\eta} k^2 P_0(k)\,c\, \frac{4\pi}{3} \int^{\Lambda} \dd q\, P_0(q) =
    - 2 \e^{4\Delta\eta} k^2 P_0(k) \,c\, \sigmad\,,
    \label{eq:P1L_h}
\end{equation}
where
\begin{equation}
    \sigmad = \frac{4\pi}{3} \int^{\Lambda} \dd q\, P_0(q)
    \label{eq:sigma_d_def}
\end{equation}
is the displacement dispersion. $P_0$ is the input linear power spectrum from
the initial condition, as defined above. From here on we will only consider the
CDM+baryon power spectrum, and we drop the cb subscript for brevity.
Due to momentum conservation, the kernels $F^{(n)}(\vq_1,\dotsc,\vq_n)$ scale
as $k^2$ when the sum $\vk = \sum_i \vq_i$ goes to zero. Therefore, the
dominant contribution to the hard limit above comes from the diagram containing
$F^{(3)}$, and the coefficient $c$ is given by
\begin{equation}
    c = 9 \int \frac{\dd \Omega_q}{4\pi}
    \left(
        \lim_{q\to\infty} \, \frac{q^2}{k^2} F^{(3)}(\vk,\vq,-\vq) F^{(1)}(\vk)
    \right)
    = \frac{61}{210}\, .
    \qquad
    \left[
        \text{1F scheme}
    \right]
    \label{eq:c_1F}
\end{equation}

We expect the $k^2$-scaling guaranteed by momentum conservation also in the 2F
scheme, but the $c$-coefficient can change and acquire a time-dependence
inherited from the kernels. To analyse the UV and determine $c(\eta)$ in the 2F
model, we compute the hard limit of the one-loop integral numerically. This can
be done as follows: Define the one-loop \emph{integrand} $\mathsf{p}_{1L}$ by
\begin{equation}
    P_{1L}(k,\eta) = \e^{4\Delta\eta} \int^{\Lambda} \dd q \, q^2 P_0(q) \int \dd \Omega_q \,
    \mathsf{p}_{1L}(\vk,\vq; \eta)\, .
    \label{eq:p_1L}
\end{equation}
It contains the different diagrams contributing to the one-loop correction with
the corresponding kernels and an additional linear power spectrum. In the hard
limit, the leading contribution to $\mathsf{p}_{1L}$ should scale as
$F^{(3)}(\vk,\vq,-\vq) \sim k^2/q^2$, therefore we can factorize the
integration:
\begin{equation}
    P_{1L}^{h}(k,\eta) =
    3 \e^{4\Delta\eta} \int \frac{\dd \Omega_q}{4\pi}
    \left(
        \lim_{q\to\infty} \, q^2 \mathsf{p}_{1L}(\vk,\vq; \eta)
    \right)
    \frac{4\pi}{3} \int^{\Lambda} \dd q P_0(q)
    \equiv
    \e^{4\Delta\eta} p_{1L}^{h}(k,\eta) \, \sigmad\,.
    \label{eq:p1L_h}
\end{equation}
To evaluate $p_{1L}^{h}$, we fix the loop momentum to a large value
$q\gg\Lambda$ and integrate over the angle numerically. Comparing to
Eq.~\eqref{eq:P1L_h}, we find
\begin{equation}
    c(\eta) = - \frac{p_{1L}^{h}(k,\eta)}{2 k^2 P_0(k)}\,.
    \label{eq:c_2F}
\end{equation}

\begin{figure}[t]
    \centering
    \includegraphics[width=0.55\linewidth]{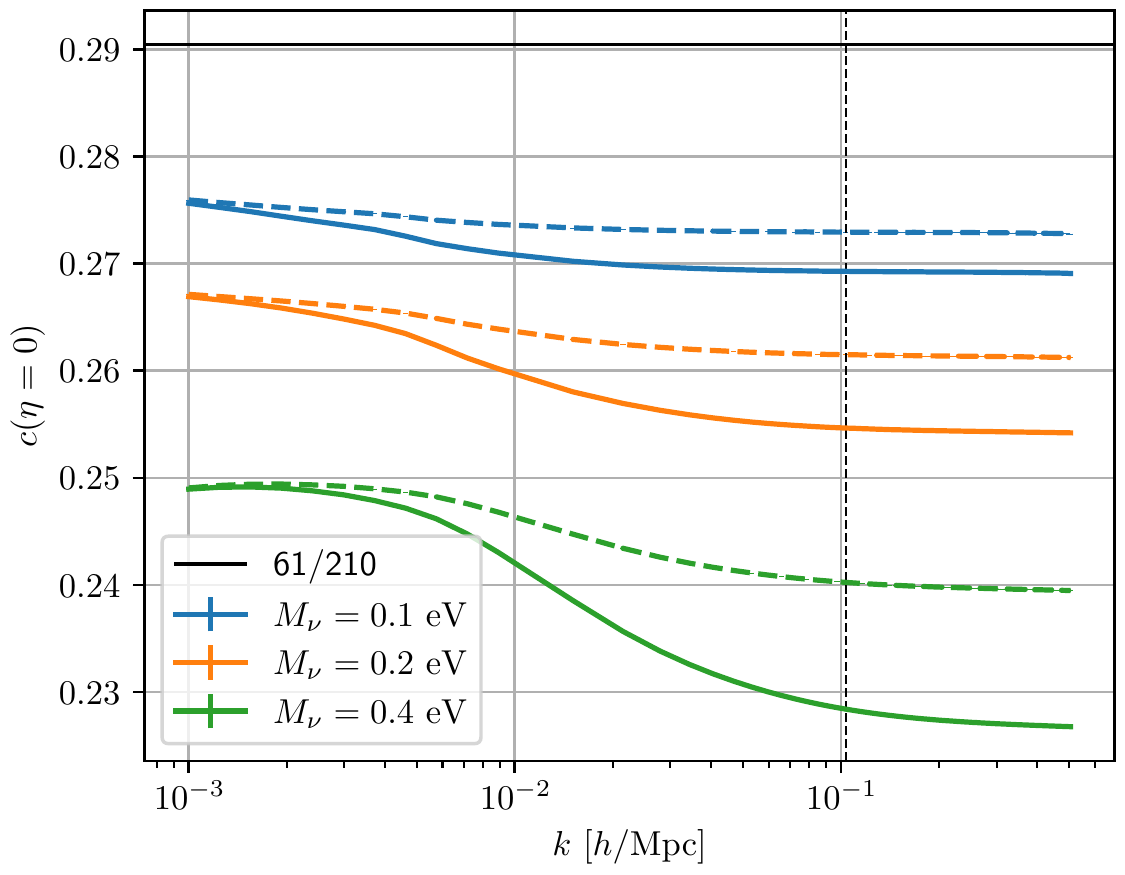}
    \caption{Hard limit of the one-loop correction in the 1F and 2F schemes
    (for the three neutrino masses), represented by the coefficient $c$ today
    ($\eta = 0$). In the dashed lines the linear kernel is divided out:
    $c/F_1^{(1)}(k)|_{\eta = 0}$. The black, dashed line indicates where we
    read off the value in each model.}
    \label{fig:c_coeff}
\end{figure}

In Fig.~\ref{fig:c_coeff} we show the results for the $c$-coefficient at
$\eta = 0$ for the different neutrino masses, using $\Lambda = 1\ihMpc$ and
fixing $q = 10\ihMpc$. The constant $c=61/210$ in the 1F scheme is shown for
comparison. Furthermore, we display dashed lines where the linear kernel is
divided out in order to isolate the scaling of the $F^{(3)}$ kernel in the hard
limit, i.e.\ $c/F_1^{(1)}(k)|_{\eta = 0}$. We see that the presence of an
additional scale, the free-streaming scale $\kFS$, induces a $k$-dependence:
$c = c(k^2/\kFS^2)$~%
\footnote{The free-streaming scale only enters the equations of motion
\eqref{eq:2F_omega} via the dimensionless ratio $k^2/\kFS^2$.}.
In the limit $k\ll \kFS$, neutrinos behave as dark matter, and the system can
be mapped onto a set of standard dark matter fluid equations for which the hard
limit yields $c$ constant. We do not recover the 1F limit of $61/210$, because
the EdS approximation is relaxed and because the kernel
$F^{(3)}(\vk,\vq,-\vq;\eta)$ is sensitive to the dynamics at $q \gg \kFS$ where
the growth rate is dampened due to neutrino free-streaming. For the largest
neutrino mass with the largest free-streaming wavenumber, this limit is visible
around $k = 10^{-3}\ihMpc$ in Fig.~\ref{fig:c_coeff}.

On scales much smaller than the free-streaming scale, neutrinos do not cluster
and the CDM+baryon and neutrino fluid system effectively decouple. The
CDM+baryon fluid is then equivalent to the standard case (1F), with the EdS
approximation relaxed and with reduced growth due to the effect of the
neutrinos on the background evolution. Hence $c$ approaches a constant for
$k\gg\kFS$.

Our aim is to obtain a renormalized power spectrum in the mildly non-linear
regime, $k\simeq 0.1\ihMpc$, therefore we assume a large scale separation
$\kFS \ll k \ll q$~%
\footnote{Note that the free-streaming scale as defined in Eq.~\eqref{eq:kFS}
is not a single scale, but rather a function of scale and time. We have
approximately $\kFS(k, z) \propto (1+z)^{-1/2}$ and a slight, monotonic
increase as a function of wavenumber.}.
In this limit $c$ is constant, which implies that the
counterterm~\eqref{eq:delta_counter_fin} can absorb the cutoff dependence of
the one-loop correction, as we show explicitly below.
For the lowest neutrino masses, we see indeed a plateau in the value of $c$
around $k = 0.1\ihMpc$ in Fig.~\ref{fig:c_coeff}. The measured value is
read off at the dashed line and quoted in Tab.~\ref{tab:EFT_params}. For the
largest neutrino mass, the assumption of large scale separation $\kFS \ll
k$ starts to break down (from Eq.~\eqref{eq:kFS} we have e.g.\
$\kFS(k = 0.1\ihMpc, \eta = 0) = 0.07 \ihMpc$), and the plateau emerges
only on even smaller scales. Nevertheless, we read off the value of $c$ and
include the model in our analysis for illustration. In practice, the limit $\kFS
\ll k \ll q$ is well satisfied for relevant values of the sum of neutrino
masses.

The counterterm~\eqref{eq:delta_counter_fin} gives the additional contribution
\begin{equation}
    P_{1L}^{\mathrm{ctr}}(k,\eta; \Lambda) =
    - 2 \e^{4\Delta\eta}\, \gamma_1(\Lambda) k^2 P_0(k)\, ,
    \label{eq:P1L_ctr}
\end{equation}
to the power spectrum, which is exactly of the form to absorb the UV
sensitivity~\eqref{eq:P1L_h}. The renormalized one-loop CDM+baryon power
spectrum therefore reads
\begin{equation}
    P_{1L}^{\mathrm{ren}}(k,\eta, \Lambda) =
    P_{1L}(k,\eta; \Lambda) + P_{1L}^{\mathrm{ctr}}(k,\eta; \Lambda)\,.
\end{equation}
Demanding that the renormalized result is cutoff-independent yields the
renormalization group equation
\begin{equation}
    0 = \frac{\dd}{\dd \Lambda}
    \left[
        P_{1L}(k,\eta; \Lambda) + P_{1L}^{\mathrm{ctr}}(k,\eta; \Lambda)
    \right]
    =
    - 2 \e^{4\Delta\eta} k^2 P_0(k)
    \frac{\dd}{\dd \Lambda}
    \left[
        c(\eta) \sigmad(\Lambda) + \gamma_1(\Lambda)
    \right]
    \label{eq:rge_L1}
\end{equation}
at leading power in gradients $k^2/\kNL^2$, which has the solution
\begin{equation}
    \gamma_1(\Lambda) =
    \overline{\gamma}_1 +
    \frac{4\pi c}{3} \int_{\Lambda}^{\infty} \dd q P_0(q)\,,
    \label{eq:gamma_1}
\end{equation}
where $\overline{\gamma}_1$ is an initial condition.

\subsection{Renormalization of the two-loop correction}
\label{subsec:L2_ren}

Next we discuss the UV-sensitivity of the two-loop correction and how we can
treat it in the EFT. The second order contributions to the effective stress
term are~\cite{Baldauf:2014qfa, Angulo:2014tfa}
\begin{equation}
    \tau_{\theta} \big|_{2} =
    d^2 \Delta \delta^{(2)}
    + e_1 \Delta \left[ \delta^{(1)} \right]^2
    + e_2 \Delta \left[ s_{ij}^{(1)} s^{(1)ij} \right]
    + e_3 \partial_i \left[ s^{(1)ij} \partial_j \delta^{(1)} \right]\,,
    \label{eq:tau_2nd_order}
\end{equation}
where the number superscript indicates the perturbative order and the tidal
tensor is defined as
\begin{equation}
    s^{ij} =
    \left(
        \partial^i\partial^j
        - \frac{1}{3} \delta^{ij} \partial^2
    \right)
    \Phi\,.
    \label{eq:s^ij}
\end{equation}
Three additional EFT parameters $e_1$, $e_2$ and $e_3$ appear at this order. In
writing down Eq.~\eqref{eq:tau_2nd_order} we have used again that in massless
neutrino cosmologies (1F) $\theta^{(1)} = - \mH f \delta^{(1)}$. In addition, the
operator $\Delta\theta^{(2)}$ can be written as a linear combination of the
existing operators, hence it is also redundant. These simplifications do not
hold in general for massive neutrino cosmologies (2F). While the linear CDM+baryon
perturbations could be related far below the free-streaming scale,
$\Delta\theta^{(2)}$ is not redundant in general. Furthermore, there are
additional operators including the neutrino perturbations at second order as
well as mixed operators with linear CDM+baryon and neutrino perturbations.
Given the smallness of the neutrino perturbations around the weakly non-linear
scales those should be negligible however.

Even in the massless case, solving the perturbative equations of motion in the
presence of the second order counterterms~\eqref{eq:tau_2nd_order} is quite
complex. Moreover, in practice, there are significant degeneracies between the
EFT parameters when calibrating to simulations. Therefore, making an ansatz for
a linear relation between the parameters and fitting the overall amplitude is
typically sufficient~\cite{Baldauf:2015aha}. One also easily overfits the data
when introducing too many free parameters (we return to this issue in
Sec.~\ref{sec:numerics}). All in all, we follow the prescription
of~\cite{Baldauf:2015aha} for renormalizing the two-loop correction and extend
it to the 2F scheme. This works well for the power spectrum and an analogous
method has also been applied successfully to the bispectrum at
two-loop~\cite{Baldauf:2021zlt}. The upshot is that we do not need to know the
specific form of the second order effective stress term, also in the massive
neutrino case.

Note that for renormalizing the two-loop, one in principle also needs
counterterms from the third order effective stress term. However, for the power
spectrum, they have been argued to yield contributions that are degenerate with those arising
at lower order~\cite{Foreman:2015lca}. The explicit third order correction to the stress
tensor is not needed within the approach followed here.

To study the UV limit of the two-loop correction, it is useful to distinguish between
the \emph{double-hard} (hh) limit where both loop momenta are hard compared to
the external momentum $k$, and the \emph{single-hard} (h) limit where one of
the momenta are hard compared to the external and the other loop momentum. We
discuss first the double-hard limit.

\subsubsection{Double-hard limit}
\label{subsubsec:hh_limit}

\begin{figure}[t]
    \centering
    \includegraphics[width=\linewidth]{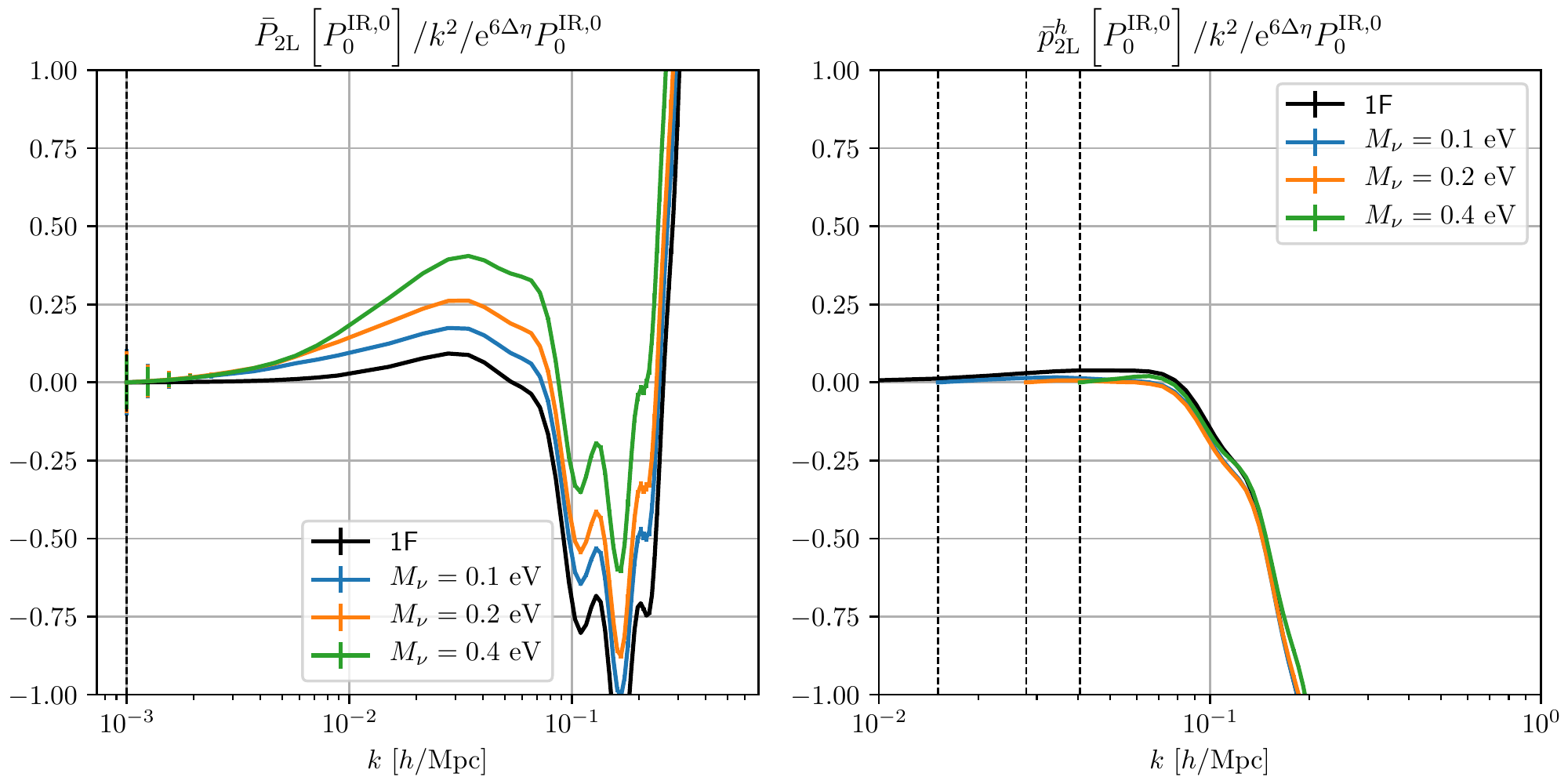}
    \caption{Subtracted two-loop and subtracted single-hard two-loop power
    spectrum correction over $k^2 P_0(k)$. We plot the corresponding IR
    resummed quantities as defined in Eq.~\eqref{eq:P2L_sub_IR_resummed} and
    \eqref{eq:p2L_h_sub_IR_resummed} below, rescaled by the two-loop growth
    factor power $\e^{6\Delta\eta}$. The three neutrino mass models computed in
    2F as well as the result in 1F are shown (for simplicity we set $\Mnu =
    0.1~\eV$ for 1F; any difference due to neutrino mass enters there only in
    the input linear power spectrum). Errorbars show uncertainty from the Monte
    Carlo integration. The dashed vertical lines indicate the subtraction
    points defining the limits $b^{hh}$ (left panel) and $b^h$ (right panel).
    The double-hard subtraction points coincide for all neutrino masses, while
    for the single-hard limit the subtraction points left to right correspond
    to increasing neutrino mass.}
    \label{fig:L2_sub}
\end{figure}

In the 1F case, the leading contribution to the double-hard region of the
two-loop comes from the propagator correction diagram containing
$F^{(5)}$~\cite{Baldauf:2015aha}. This follows from the $k^2$-scaling of the
kernels in the limit $k/q_1 \propto k/q_2 \ll 1$, and also holds in the
2F case guaranteed by momentum conservation. The other diagrams enter in this
limit with additional powers of $k^2$ and can be neglected at leading order in
gradients. The $F^{(5)}$-diagram contribution scales as $k^2 P_0(k)$ and is
therefore degenerate with the UV-limit at one-loop and can be absorbed by the
counterterm~\eqref{eq:P1L_ctr}. We can split the corresponding EFT parameter at
NNLO into a one- and two-loop part,
\begin{equation}
    \gamma_1^{\mathrm{NNLO}} = \gamma_1^{1L} + \gamma_1^{2L}.
\end{equation}
Following~\cite{Baldauf:2015aha}, we use a renormalization scheme where we
choose $\gamma_1^{2L}$ so that it exactly cancels the double-hard region of the
two-loop, while $\gamma_1^{1L}$ is calibrated to simulations and contains the
finite part of the counterterm. We drop the superscript and use
$\gamma_1 \equiv \gamma_1^{1L}$ in the numerical analysis below. This means in
practice that we subtract the double-hard contribution from the two-loop
correction:
\begin{equation}
    \bar{P}_{2L}(k,\eta; \Lambda) \equiv P_{2L}(k,\eta; \Lambda)
        - P_{2L}^{hh}(k,\eta; \Lambda) \,.
    \label{eq:P2L_sub}
\end{equation}
The double-hard contribution $P_{2L}^{hh}$ can be computed analytically with
EdS-kernels in the 1F scheme (see e.g.~\cite{Baldauf:2015aha}), and numerically
in the massive neutrino models by fixing the loop momenta to large values.
Alternatively, one can fit the low-$k$ region by computing
$b^{hh} = \lim_{k\to 0} P_{2L}/k^2/P_0$ and subtracting
$P_{2L}^{hh} = b^{hh} k^2 P_0$. The double-hard limit is realized
in this region because the loop integral has sole support from wavenumbers
$q_1,q_2 \gg k$. We use the latter method. In the left panel of
Fig.~\ref{fig:L2_sub} we show the ratio $\bar{P}_{2L}/k^2/P_0$ for the 1F
scheme and for the different neutrino masses using the 2F scheme (more
precisely, we show the corresponding IR resummed quantities defined in
Eqs.~\eqref{eq:P2L_sub_IR_resummed} and \eqref{eq:p2L_h_sub_IR_resummed}
below). The results are obtained using a cutoff $\Lambda = 1\ihMpc$. In the 1F
scheme, we see that the two-loop attain the $k^2P_0(k)$-scaling below $k\simeq
10^{-2} \ihMpc$. Due to the presence of the free-streaming scale in the
2F neutrino models, they only recover the asymptotic $k^2 P(k)$-scaling for
even smaller $k$. We read off the limit $b^{hh}$ at $k = 10^{-3}\ihMpc$ (dashed
vertical line in Fig.~\ref{fig:L2_sub}). Note that $b^{hh}$ defines the
renormalization point, and changing its value only leads to a shift in the EFT
parameter $\gamma_1$. We check that adjusting the wavenumber at which we read
off the limit $b^{hh}$ has insignificant impact on our conclusions, up to the
shift. Therefore we find it convenient to evaluate the limit for $k \ll \kFS$,
while we will obtain the corresponding limit $b^{h}$ for the single-hard two-loop
contribution at $k \gg \kFS$ as we discuss below.%
\footnote{One could obtain $b^{hh}$ at $k\gg\kFS$ numerically by
fixing the loop momenta to a large value $q_1,q_2 \gg k$, yielding a different
value than on scales larger than the free-streaming length, however ultimately
this would just lead to a shift in the measured value of $\gamma_1$.}

\subsubsection{Single-hard limit}
\label{subsubsec:h_limit}

Next, we consider the limit in which one of the loop momenta becomes large:
$q_1 \gg k$, $q_2 \sim k$ (or equivalently with $q_1$, $q_2$ switched). The
counterterms that renormalize this limit are one-loop diagrams with insertions
of EFT operators. Computing those diagrams is rather complicated even in the 1F
case, and would in the 2F scheme potentially involve further operators from the
generalization of \eqref{eq:tau_2nd_order} including neutrino perturbations.
Therefore we opt for the approach put forward in~\cite{Baldauf:2015aha}, using
an ansatz for the single-hard counterterm with one extra EFT parameter.

Analogously to the one-loop case, we start by defining the two-loop
\emph{integrand} $\mathsf{p}_{2L}$:
\begin{equation}
    P_{2L}(k,\eta) = \e^{6\Delta\eta}
    \int^{\Lambda} \dd q_1 \, q_1^2 P_0(q)
    \int^{\Lambda} \dd q_2 \, q_2^2 P_0(q)
    \int \dd \Omega_{q_1} \dd \Omega_{q_2}\,
    \mathsf{p}_{2L}(\vk,\vq_1,\vq_2; \eta)\, .
    \label{eq:p_2L}
\end{equation}
In the single-hard limit $q_1 \to\infty$, the leading contribution comes from
kernels $F^{(n)}(\dotsc,\vq_1,-\vq_1,\dotsc) \propto 1/q_1^2$. The two-loop
integral can thus be factorized in this limit giving the single-hard result
\begin{equation}
    P_{2L}^{h}(k,\eta) =
    2 \e^{6\Delta\eta} \, p_{2L}^{h}(k,\eta) \, \sigmad\,,
    \label{eq:P2L_h}
\end{equation}
where
\begin{equation}
    p_{2L}^{h}(k,\eta) \equiv
    3 \int^{\Lambda} \dd q_2 \, q_2^2 P_0(q)
    \int \frac{\dd \Omega_q}{4\pi}
    \left(
        \lim_{q_1\to\infty} \, q_1^2 \mathsf{p}_{2L}(\vk,\vq_1,\vq_2; \eta)
    \right)\,.
    \label{eq:p2L_h}
\end{equation}
The factor $2$ in \eqref{eq:P2L_h} accounts for the equivalent contribution
where $q_2$ is hard. We compute the limit \eqref{eq:p2L_h} both in the 1F and
2F schemes numerically by fixing the hard loop momentum to a large value $q_1
\gg \Lambda$ and evaluating the resulting integral using Monte Carlo
integration.

Following the prescription of~\cite{Baldauf:2015aha}, we assume that we can
correct for the spurious UV contribution in the single-hard limit by a shift in
the value of the displacement dispersion $\sigmad$, i.e.\
\begin{equation}
    \sigmad(\Lambda) \mapsto \sigmad(\Lambda) + \mathcal{N}\gamma_2(\Lambda)\,,
    \label{eq:sigmad_shift}
\end{equation}
where the $\gamma_2$-term corresponds to the EFT correction and $\mathcal{N}$ is
a factor introduced for convenience that we define below. The replacement yields
an additional counterterm
\begin{equation}
    2 \e^{6\Delta\eta} \, p_{2L}^{h}(k,\eta) \, \mathcal{N} \gamma_2\,,
    \label{eq:p2L_ctr}
\end{equation}
to the power spectrum. This term could in principle immediately be added to the
renormalized power spectrum, however we notice that part of it is still
degenerate with the double-hard limit. The definition \eqref{eq:p2L_h}
integrates over a hard region $q_2 \gg k$ for external wavenumbers far below
the cutoff, which yields a contribution proportional to $k^2 P_0(k)$. Such a
contribution is absorbed by the $\gamma_1$-counterterm defined above, however
as discussed for the double-hard limit, we adopt a renormalization scheme where
$\gamma_1^{2L}$ exactly cancels the double-hard region of the two-loop, now
including also the double-hard contribution from the counterterm
\eqref{eq:p2L_ctr}. In practice this means that we subtract the degenerate
double-hard part from the single-hard contribution in analogy to the subtracted
two-loop correction:
\begin{equation}
    \bar{p}_{2L}^{h} = p_{2L}^{h} - p_{2L}^{hh}\,.
    \label{eq:p2L_h_sub}
\end{equation}

To obtain the double-hard limit $p_{2L}^{hh}$, we use the same method as for
the full two-loop correction: we evaluate numerically the limit
$b^{h} = \lim_{k\to 0} p_{2L}^{h} / k^2 / P_0(k)$. In this limit the
integral~\eqref{eq:p2L_h} only has support where $q_2 \gg k$ and since $q_1$ is
hard we recovered the double-hard limit and $p_{2L}^{hh} = b^{h} k^2 P_0(k)$.

The subtracted single-hard limit $\bar{p}_{2L}^{h}$ divided by the double-hard
scaling $k^2 P_0(k)$ is shown in the right panel of Fig.~\ref{fig:L2_sub} (we
plot the corresponding IR resummed quantities defined in the next section). We
set the cutoff to $\Lambda = 1\ihMpc$ and the hard loop momentum to
$q_1 = 10\ihMpc$. In the 1F scheme, the low-$k$ limit has indeed the $k^2 P_0$
scaling, and we subtract the value at $k=10^{-2}\ihMpc$ so that the single-hard
limit $\bar{p}_{2L}^{h}$ does not give a contribution degenerate with the
one-loop counterterm. For the 2F scheme we also find the $k^2 P_0$ scaling is approached,
however for very small wavenumbers certain deviations occur. They may in part be attributed
to large numerical cancellations due to our choice $q_1=10 \ihMpc$, and we checked that
the $k^2 P_0$ scaling extends to lower $k$ for $q_1=5 \ihMpc$, without changing the result within
the relevant range $k\gtrsim 0.05 \ihMpc$ where $\bar{p}_{2L}^{h} $ gives a significant
contribution to the renormalized two-loop power spectrum. We therefore use a subtraction
point $k\simeq 0.015, 0.027, 0.04\ihMpc$ for $M_\nu=0.1,0.2,0.4~\eV$, to read off $b^{h}$ and subtract the double-hard
contribution (indicated by the vertical dashed lines in Fig.~\ref{fig:L2_sub}).  As for the full two-loop correction, the
definition of the double-hard subtraction point and corresponding value $b^{h}$
determines the renormalization point, hence a change of $b^{h}$ only results in
a shift of $\gamma_1$. We check in particular that choosing $0.027\ihMpc$ as
subtraction points also for $\Mnu = 0.1$ and $0.4~\eV$ has negligible impacts
on the results.

All in all, we have the two-loop counterterm
\begin{equation}
    P_{2L}^{\mathrm{ctr}}(k,\eta; \Lambda) =
    2 \e^{6\Delta\eta} \mathcal{N} \gamma_2(\Lambda)\, \bar{p}_{2L}^{h}(k,\eta)\,,
    \label{eq:P2L_ctr}
\end{equation}
and the renormalized two-loop correction reads
\begin{equation}
    P_{2L}^{\mathrm{ren}}(k,\eta; \Lambda) =
    \bar{P}_{2L}(k,\eta; \Lambda) + P_{2L}^{\mathrm{ctr}}(k,\eta; \Lambda)\,.
    \label{eq:P2L_ren}
\end{equation}
After renormalization, the CDM+baryon power spectrum at NNLO should be
independent of the cutoff:
\begin{equation}
    \frac{\dd}{\dd \Lambda} P_{\mathrm{NNLO}}^{\mathrm{ren}}(k,\eta;\Lambda) = 0\,,
    \label{eq:rge_nnlo}
\end{equation}
where
\begin{equation}
    P_{\mathrm{NNLO}}^{\mathrm{ren}}(k,\eta;\Lambda) =
    P_{\mathrm{tree}}(k,\eta) +
    P_{1L}^{\mathrm{ren}}(k,\eta; \Lambda) +
    P_{2L}^{\mathrm{ren}}(k,\eta; \Lambda)\,.
    \label{eq:P_nnlo}
\end{equation}
By removing the double-hard contribution to the two-loop correction (i.e.\ by
using the subtracted quantities $\bar{P}_{2L}$ and $\bar{p}_{2L}^{h}$), there
are no degenerate UV contributions between the one- and two-loop corrections,
and we can solve the RGE~\eqref{eq:rge_nnlo} independently for each term.
Furthermore, the one-loop RGE~\eqref{eq:rge_L1} and its solution stay the
same.

The remaining cutoff-dependence comes from the single-hard region of the
two-loop correction, which should be corrected for by the two-loop counterterm,
\begin{equation}
    0 = \frac{\dd}{\dd \Lambda}
    \left[
        \bar{P}_{2L}^{h}(k,\eta;\Lambda) +
        P_{2L}^{\mathrm{ctr}}(k,\eta; \Lambda)
    \right]
    =
    2 \e^{6\Delta\eta} \bar{p}_{2L}^{h}(k,\eta)
    \frac{\dd}{\dd \Lambda}
    \left[
        \sigmad(\Lambda) + \mathcal{N}\gamma_2(\Lambda)
    \right]\,,
    \label{eq:rge_L2}
\end{equation}
valid at leading power in gradients. This result reflects the assumption that
the single-hard region is regulated by a shift in the displacement dispersion
$\sigmad$. The solution is
\begin{equation}
    \gamma_2(\Lambda) =
    \overline{\gamma}_2 +
    \frac{4\pi}{3\mathcal{N}} \int_{\Lambda}^{\infty} \dd q\, P_0(q)\,,
    \label{eq:gamma_2}
\end{equation}
where $\overline{\gamma}_2$ is an initial condition.%
\footnote{For the EFT parameters $\gamma_1$ and $\gamma_2$ we use the bar
notation to indicate the value as $\Lambda\to\infty$. This is unrelated to the
bar notation we adopt for the \emph{subtracted} power spectra, e.g.\
$\bar{P}_{2L}$ as defined in Eq.~\eqref{eq:P2L_sub}.}
We choose $\mathcal{N} = 1/c$, with $c$ defined in Eq.~\eqref{eq:c_1F} (and
evaluated in the 2F scheme using Eq.~\eqref{eq:c_2F}), in order to treat $\gamma_1$
and $\gamma_2$ on equal footing (cf.\ RGE for $\gamma_1$~\eqref{eq:rge_L1}).

\begin{figure}[t]
    \centering
    \includegraphics[width=0.95\textwidth]{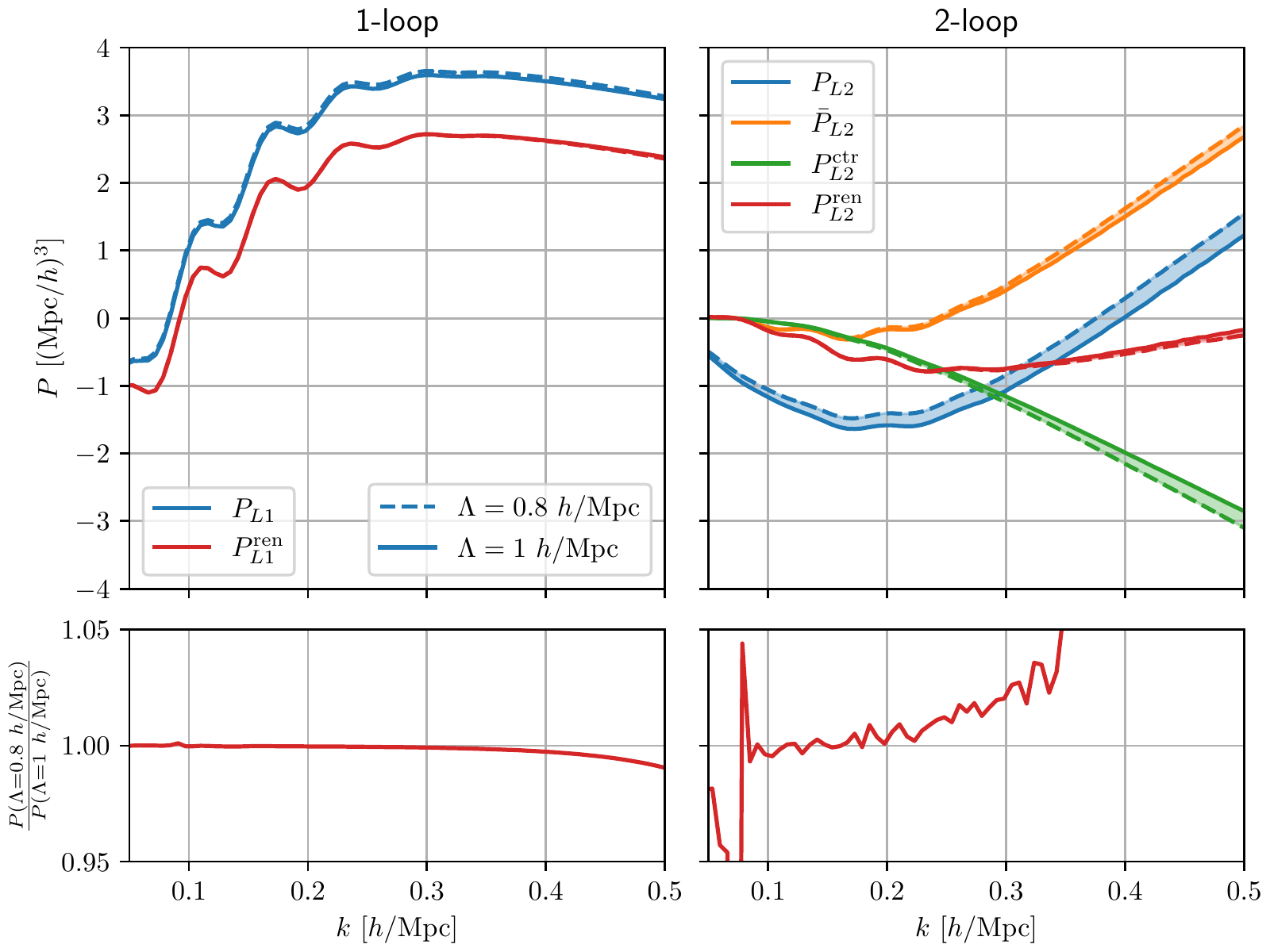}
    \caption{Cutoff-dependence of one- and two-loop corrections. Dashed and
    solid lines display results using $\Lambda = 0.8~h/\Mpc$ and $\Lambda =
    1~h/\Mpc$, respectively, and the colored bands show the difference.
    \emph{Upper left:} bare and renormalized result at one-loop. \emph{Upper
    right:} bare, subtracted and renormalized two-loop correction, as well as
    the two-loop counterterm. The fractional difference between the
    renormalized power spectra using the two cutoffs are shown in the lower
    panel. The EFT parameters are set to $\gamma_1 = \gamma_2 = 1~\Mpc^2/h^2$.}
    \label{fig:cutoff_dependence}
\end{figure}

To check that we indeed obtain cutoff-independent loop corrections, we show the
various loop and counterterm results in Fig.~\ref{fig:cutoff_dependence}, for
$\Mnu = 0.1~\eV$ using two cutoffs $\Lambda = 0.8$ (dashes lines) and $1\ihMpc$
(solid lines). The top panels show the absolute contributions, while the bottom
panels show the fractional difference between the renormalized power spectra
using the two cutoffs. We see that the cutoff-dependence of the renormalized
one-loop correction is on the permille level up to $k\simeq 0.4\ihMpc$; for
increasing $k$ the neglected next-to-leading gradient corrections
$\mathcal{O}(k^4/\kNL^4)$ become more and more important. The double-hard limit
gives the largest contribution to the bare two-loop correction, as indicated by
the bare (blue) and subtracted (yellow) graphs in the right panel of
Fig.~\ref{fig:cutoff_dependence}. The single-hard limit is regulated by the
counterterm (green) yielding the renormalized two-loop correction (red). We
display its fractional difference using the two cutoffs in the bottom plot,
where the large \emph{relative} deviations at small $k$ come from the
renormalized correction being very close to zero. The difference is
less than 1\% up to about $k=0.25\ihMpc$.

\subsection{IR resummation}

A well-known issue with the Eulerian perturbative description is that the
effect of large bulk flows on the BAO peak is inadequately
modelled~\cite{Eisenstein:2006nj,Crocce:2007dt}. Nevertheless, this effect can
relatively straightforwardly be taken into account non-perturbatively by
resumming the contributions from long-wavelength displacements, known as IR
resummation~\cite{Senatore:2014via,Baldauf:2015xfa,Vlah:2015zda,Blas:2016sfa}.
We use IR resummation for both schemes 1F and 2F in order to model the BAO
scales as accurately as possible.

The large bulk flows affect only the BAO wiggles and not the broadband part of
the power spectrum, therefore it is useful to split the linear power spectrum
into a wiggly and non-wiggly part
\begin{equation}
    P_0(k) = \Pnw(k) + \Pw(k)\,.
    \label{eq:wiggly_split}
\end{equation}
We perform this split using the method described
in~\cite{Hamann:2010pw,Chudaykin:2020aoj}, i.e.\ we obtain $\Pnw$ by Fourier
transforming the power spectrum to real space, removing the BAO peak, and
smoothly interpolating the power spectrum without the peak.
Following~\cite{Blas:2016sfa}, the next step is to compute the damping factor
\begin{equation}
    \Sigma^2 = \frac{4\pi}{3} \e^{2\Delta\eta}\int_{0}^{k_s} \dd q\, \Pnw(q)
    \left[
    1 - j_0
    \left(\frac{q}{\kosc}\right)
    + 2 j_2
    \left(\frac{q}{\kosc}\right)
    \right]\, ,
    \label{eq:Sigma2}
\end{equation}
where $j_n$ are spherical Bessel functions, $\kosc = h/(110~\Mpc)$ is the
wavenumber that corresponds to the BAO period and $k_s$ is the separation scale
of long and short modes, introduced to have separate perturbative expansions in
the two regimes. We use $k_s = 0.2\ihMpc$, as recommended
by~\cite{Blas:2016sfa}.%
\footnote{We checked that changing the separation scale to $k_s = 0.1\ihMpc$
has negligible impacts on our results and conclusions.}
The IR resummed power spectrum at $N$-th order is
\begin{equation}
    P^{\IR} =
    \sum_{L=0}^{N} P_{L\text{-loop}}
    \left[
        \Pnw + \Pw \, \e^{-k^2 \Sigma^2}
        \sum_{n=0}^{N-L} \frac{(k^2 \Sigma^2)^n}{n!}
    \right]
    \equiv
    \sum_{L=0}^{N} P_{L\text{-loop}}
    \left[
        P_0^{\IR,N-L}
    \right]\,,
    \label{eq:P_ir_resum}
\end{equation}
where the bracket notation $P_{L\text{-loop}}[X]$ indicates that $X$ is the
linear input power spectrum that enters in the $L$-loop integrals. The
exponential damping factor resums the infrared contributions, while the sum
over $n$ corrects for overcounting (the IR contributions are also included
perturbatively order by order in the loop evaluations)~\cite{Blas:2016sfa}.%
\footnote{We have $\lim_{N\to\infty} P_0^{N-L} = \Pnw + \Pw = P_0$ such that
the exact spectra with and without IR resummation agree; the difference arises
only when working at finite order in perturbation theory.}
In the last equality we defined a shorthand notation for the input power
spectrum in the IR resummed result $P_0^{\IR,N-L}$, i.e.\ the quantity in the
bracket, which depends on the difference $N-L$. Specifically, at the orders in
perturbation theory in which we work one has
\begin{align}
    P_{\mathrm{LO}}^{\mathrm{IR}} &= P_{\mathrm{tree}}
    \left[
        \Pnw(k) + \e^{-k^2 \Sigma^2} \Pw(k)
    \right]\,,
\\
    P_{\mathrm{NLO}}^{\mathrm{IR}} &= P_{\mathrm{tree}}
    \left[
        \Pnw(k) + \e^{-k^2 \Sigma^2} ( 1 + k^2\Sigma^2) \Pw(k)
    \right]
    \nonumber\\
    & \phantom{=} + P_{1L}
    \left[
        \Pnw(k) + \e^{-k^2 \Sigma^2} \Pw(k)
    \right]\,,
\\
    P_{\mathrm{NNLO}}^{\mathrm{IR}} &= P_{\mathrm{tree}}
    \left[
        \Pnw(k) + \e^{-k^2 \Sigma^2}
        \left(
        1 + k^2\Sigma^2 + \frac{1}{2} (k^2\Sigma^2)^2
        \right)
        P_w(k)
    \right]
    \nonumber \\
    & \phantom{=} + P_{1L}
    \left[
        \Pnw(k) + \e^{-k^2 \Sigma^2} ( 1 + k^2\Sigma^2) \Pw(k)
    \right]
    \nonumber \\
    & \phantom{=} + P_{2L}
    \left[
        \Pnw(k) + \e^{-k^2 \Sigma^2} \Pw(k)
    \right]\,.
\end{align}
Note that the tree-level power spectrum is non-trivial in the 2F scheme and
given by
\begin{equation}
    P_{\mathrm{tree}}\left[ P_0 \right](k,\eta)
    = \left[ F^{(1)}(k,\eta) \right]^2 P_0(k)\,.
    \label{eq:P_tree}
\end{equation}

In deriving Eq.~\eqref{eq:P_ir_resum} in the Eulerian theory, the following
property of the kernels in the soft limit is
key~\cite{Blas:2016sfa}:
\begin{equation}
    F^{(n)}(\vq_1,\dotsc,\vq_n) \to
    \frac{m!}{n!} F^{(m)}(\vq_1,\dotsc,\vq_m)
    \frac{\vq\cdot\vq_{m+1}}{q_{m+1}^2}
    \dotsb
    \frac{\vq\cdot\vq_{n}}{q_{n}^2}
    \label{eq:Fn_soft_limit}
\end{equation}
in the limit where $q_1,\dotsc,q_m$ are much smaller than the remaining
arguments $q_{m+1},\dotsc,q_n$ and where $\vq = \sum \vq_i$. This
factorization follows from Galilean invariance~\cite{Sugiyama:2013pwa}, which
is certainly also a symmetry in the 2F scheme. Therefore the IR resummed
equation \eqref{eq:P_ir_resum} is valid also in the 2F case.

In practice, we use the following simplified expression to evaluate IR resummed
loops
\begin{multline}
    P_{L\text{-loop}}
    \left[
        \Pnw + \Pw \, \e^{-k^2 \Sigma^2}
        \sum_{n=0}^{N-L} \frac{(k^2 \Sigma^2)^n}{n!}
    \right]
    \\
    \to P_{L\text{-loop}}
    \left[
        \Pnw
    \right]
    +
    \left(
    P_{L\text{-loop}} \left[ \Pnw + \Pw \right]
    - P_{L\text{-loop}} \left[ \Pnw \right]
    \right) \times
    \e^{-k^2 \Sigma^2}
    \sum_{n=0}^{N-L} \frac{(k^2 \Sigma^2)^n}{n!}\,,
    \label{eq:P_ir_resum_alt}
\end{multline}
which is valid up to corrections of $\mathcal{O}(\Pw^2)$ as well as diagrams
involving $\Pw$ inside a hard loop. Neglecting these corrections is a good
approximation because $\Pw$ is small compared to $\Pnw$, and it oscillates
around zero so that its integral vanishes.

\subsection{All together: summary}

We end this section by writing down the final expressions for the power
spectrum up to NNLO including both IR resummation and EFT corrections. We have
\begin{align}
    P_{\mathrm{LO}}^{\mathrm{IR}}(k) &=
    P_{\mathrm{tree}}
    \left[ P_0^{\IR,0} \right]\,,
\\
    P_{\mathrm{NLO}}^{\mathrm{IR}}(k) &=
    P_{\mathrm{tree}}
    \left[ P_0^{\IR,1} \right]
    + P_{1L}
    \left[ P_0^{\IR,0} \right]
    + P_{1L}^{\mathrm{ctr}}
    \left[ P_0^{\IR,0} \right]\,,
    \label{eq:P_NLO_IR}
\\
    P_{\mathrm{NNLO}}^{\mathrm{IR}}(k) &=
    P_{\mathrm{tree}}
    \left[ P_0^{\IR,2} \right]
    + P_{1L}
    \left[ P_0^{\IR,1} \right]
    + P_{1L}^{\mathrm{ctr}}
    \left[ P_0^{\IR,1} \right]
    + \bar{P}_{2L}
    \left[ P_0^{\IR,0} \right]
    + P_{2L}^{\mathrm{ctr}}
    \left[ P_0^{\IR,0} \right]\,,
    \label{eq:P_NNLO_IR}
\end{align}
where the counterterms are given by
\begin{align}
    P_{1L}^{\mathrm{ctr}}
    \left[ P_0^{\IR,N-L} \right](k, \eta; \Lambda) & =
    - 2 \e^{4\Delta\eta}\, \gamma_1(\Lambda) k^2 P_0^{\IR,N-L}(k) \,,
    \\
    P_{2L}^{\mathrm{ctr}}
    \left[ P_0^{\IR,N-L} \right](k, \eta; \Lambda) & =
    2 \frac{\e^{6\Delta\eta}}{c(\eta)} \gamma_2(\Lambda)\, \bar{p}_{2L}^{h}
    \left[ P_0^{\IR,N-L} \right](k, \eta; \Lambda) \,,
    \label{eq:P_ctr_IR_resummed}
\end{align}
with $c$ defined in Eq.~\eqref{eq:c_1F} and the subtracted quantities
defined as
\begin{align}
    \bar{P}_{2L}
    \left[ P_0^{\IR,N-L} \right](k, \eta; \Lambda) & =
    P_{2L}
    \left[ P_0^{\IR,N-L} \right](k, \eta; \Lambda)
    - b^{hh} k^2 P_0^{\IR,N-L}(k) \,,
    \label{eq:P2L_sub_IR_resummed}
    \\
    \bar{p}_{2L}^{h}
    \left[ P_0^{\IR,N-L} \right](k, \eta; \Lambda) & =
    \bar{p}_{2L}^{h}
    \left[ P_0^{\IR,N-L} \right](k, \eta; \Lambda)
    - b^{h} k^2 P_0^{\IR,N-L}(k) \,.
    \label{eq:p2L_h_sub_IR_resummed}
\end{align}
The single-hard contribution $p_{2L}^{h}$ was defined in Eq.~\eqref{eq:p2L_h}
and the RGEs for the EFT parameters were given in Eqs.~\eqref{eq:rge_L1} and
\eqref{eq:rge_L2}. The double-hard limit constants $b^{hh}$ and $b^{h}$ are
computed in the same way as in subsections~\ref{subsubsec:hh_limit} and
\ref{subsubsec:h_limit}, but using $P_0^{\IR,N-L}$ as the linear power
spectrum, e.g\
$b^{hh} = \lim_{k\to 0} P_{2L}\left[P_0^{\IR,N-2}\right]/k^2/P_0^{\IR,N-2}$.
In general, the EFT discussion in subsections~\ref{subsec:L1_ren} and
\ref{subsec:L2_ren} above can immediately be promoted to include IR resummation
by the replacement $P_0 \to P_0^{\IR,N-L}$.

For completeness, we note again that in the 1F scheme the time-dependence is
factored out, so to obtain the power spectrum at $z = 0$ we can use $\etaini =
0$ and $\Delta\eta = 0$. In the 2F scheme on the other hand, we chose the
matching between the Boltzmann hierarchy and the two-fluid description at
$\etaini$ corresponding to $z = 25$. Therefore, to obtain the quantities
$P_0^{\IR,N-L}$, we take the linear power spectrum from CLASS at $z = 0$ or $z
= 25$ depending on the scheme, split it into its wiggly and non-wiggly
components, compute the damping factor $\Sigma^2$ and use the definition in
Eq.~\eqref{eq:P_ir_resum}.

\section{Comparison to N-body simulations}
\label{sec:numerics}

To calibrate the EFT parameters and perform a comparison of the 1F and 2F
schemes in the effective theory, we utilise the Quijote N-body
simulations~\cite{Villaescusa-Navarro:2019bje}. In particular, we use the set
of massive neutrinos simulations with $512^3$ CDM particles and $512^3$
neutrino particles in a cubic box of size $(1~\mathrm{Gpc}/h)^3$. Moreover, we
use the simulations with pair fixed initial conditions that significantly
reduce the cosmic variance for the power spectrum. The cosmological parameters are
$\Omega_m = 0.3175$, $\Omega_b = 0.049$, $h = 0.6711$, $n_s = 0.9624$,
$\sigma_{8} = 0.834$ and the neutrinos mass is $\Mnu = 0.1$, $0.2$ and
$0.4~\eV$ in the different models respectively. We estimate the power spectrum
and uncertainty at $z = 0$ using the average and standard deviation of the 500
realizations.

On the largest scales, the binning method used to obtain the power spectrum
from the simulations become significant and has to be taken into account when
comparing simulation and theory. We use the Pylians library%
\footnote{\href{https://github.com/franciscovillaescusa/Pylians}{github.com/franciscovillaescusa/Pylians}}
to perform the same binning on the linear power spectrum. Instead of binning
also the loop corrections, we can equivalently compare the theoretical results
to the ``unbinned'' N-body power spectrum
\begin{equation}
    P_{\mathrm{data}}(k) =
    \frac{P_0(k, z=0)}{P_0^{\mathrm{binned}}(k,z=0)} P_{\mathrm{data}}^{\mathrm{binned}}(k)
    \, .
    \label{eq:unbinning}
\end{equation}

The EFT parameters are calibrated by minimizing the following $\chi^2$ at NLO
and NNLO:
\begin{align}
    \chi_{\mathrm{NLO}}^2 &=
    \sum_{k = \kmin}^{\kmax}
    \frac{
        \left[
            P_{\mathrm{data}}(k) - P_{\mathrm{NLO}}^{\IR}(k,\gamma_1)
        \right]^2
    }{
    \left[
        \Delta P_{\mathrm{data}}(k)
    \right]^2
    }\,,
    \label{eq:chi2_nlo}
    \\
    \chi_{\mathrm{NNLO}}^2 &=
    \sum_{k = \kmin}^{\kmax}
    \frac{
        \left[
            P_{\mathrm{data}}(k) - P_{\mathrm{NNLO}}^{\IR}(k,\gamma_1,\gamma_2)
        \right]^2
    }{
    \left[
        \Delta P_{\mathrm{data}}(k)
    \right]^2
    }\,,
    \label{eq:chi2_nnlo}
\end{align}
where $\Delta P_{\mathrm{data}}(k)$ is the estimated uncertainty of the N-body
power spectrum. We have $k_{\mathrm{min}} = 0.0089\ihMpc$ and at
$k_{\mathrm{max}} = 0.1$ and $0.3\ihMpc$ we have 14 and 46 wavenumber grid
points in the sum, respectively. The perturbative results
$P_{\mathrm{NLO}}^{\IR}$ and $P_{\mathrm{NNLO}}^{\IR}$ are calculated using
Eqs.~\eqref{eq:P_NLO_IR} and \eqref{eq:P_NNLO_IR} with cutoff $\Lambda =
1\ihMpc$ and where the hard limits are evaluated by fixing the hard momenta to
$q_1 = 10\ihMpc$ as described in the previous section.

We consider the following cases
\begin{align*}
    &\text{NLO}                                                  && \{\bar{\gamma}_1\}                  && \text{one-loop, 1-parameter,}                                                  \\
    &\text{NLO},~\bar{\gamma}_1 = \bar{\gamma}_1^{\text{[NNLO]}} && \{\emptyset\}                       && \text{one-loop, 0-parameter with } \bar{\gamma}_1 \text{ fixed from NNLO fit,} \\
    &\text{NNLO},\bar{\gamma}_2 = \bar{\gamma}_1                 && \{\bar{\gamma}_1 \}                 && \text{two-loop, 1-parameter with } \bar{\gamma}_2 = \bar{\gamma}_1\,,          \\
    &\text{NNLO}                                                 && \{\bar{\gamma}_1, \bar{\gamma}_2 \} && \text{two-loop, 2-parameter,}                                                  \\
\end{align*}
where the NLO cases are fit using Eq.~\eqref{eq:chi2_nlo} and NNLO cases using
Eq.~\eqref{eq:chi2_nnlo}. We calibrate the EFT parameters in the limit
$\Lambda\to\infty$ (cf. Eqs.~\eqref{eq:gamma_1} and \eqref{eq:gamma_2}). The
second case, NLO $\bar{\gamma}_1 = \bar{\gamma}_1^{\text{[NNLO]}}$, is not fitted
since $\bar{\gamma}_1$ is fixed from the two-parameter NNLO case. We explain
the motivation for this case below. The third case, NNLO
$\bar{\gamma}_2 = \bar{\gamma}_1$, is motivated by the assumption that in the
EFT, the displacement dispersion is corrected in an universal manner as
$\sigmad \to \sigmad + \gamma_1/c$ both at one- and two-loop.

\subsection{Results}

\begin{figure}[t]
    \centering
    \includegraphics[width=\linewidth]{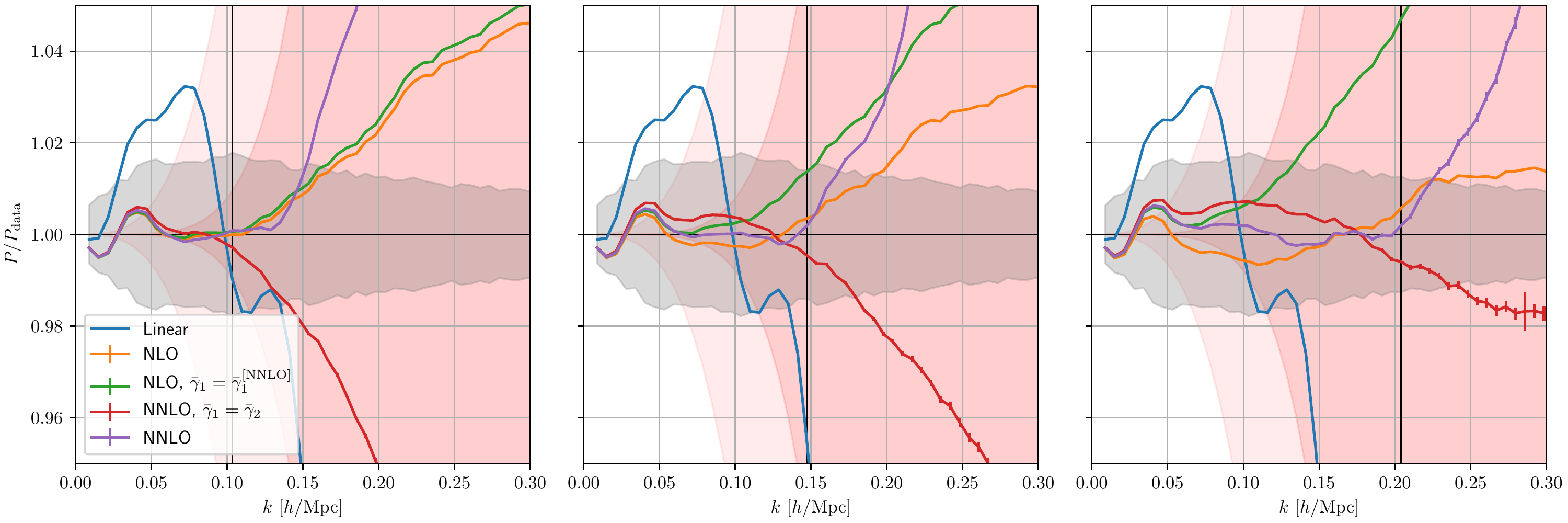}
    \caption{CDM+baryon power spectrum computed in the 2F scheme normalized to
    N-body data. We show the perturbative results at NLO and NNLO with
    different assumptions on the EFT parameters, for pivot scales $\kmax =
    0.103$ (left) $0.148$ (middle) and $0.204\ihMpc$ (right). The gray band
    indicates the N-body uncertainty, while the red shaded area display
    estimated theoretical uncertainty at one-loop (light shading) and two-loop
    (darker shading). The sum of neutrino masses is $\Mnu = 0.1~\eV$.}
    \label{fig:pk_order}
\end{figure}

We begin by comparing the full 2F solution for the various NLO/NNLO calibration
cases above with the most phenomenologically relevant neutrino mass $\Mnu =
0.1~\eV$. In Fig.~\ref{fig:pk_order} we show the power spectrum using three
pivot scales $\kmax = 0.103$, $0.148$ and $0.204\ihMpc$, normalized to the
N-body results. The gray band indicates estimated uncertainty from the N-body
simulation and the red shading corresponds to an ansatz for the expected
theoretical uncertainty at NLO (light shading) and NNLO (darker
shading)~\cite{Baldauf:2016sjb}. In all panels, there is a ``bump'' feature in
the relative difference around $k = 0.01$ -- $0.05\ihMpc$ of a few permille
(i.e.\ well within the uncertainty of the N-body data) that arises due to
finite N-body box size and binning uncertainty not captured by the
correction~\eqref{eq:unbinning}. Nevertheless, all cases can be fitted to lie
within the N-body uncertainty up to even the largest pivot scale. Depending on
the loop order, we are subject to overfitting above a certain $\kmax$. In
particular, given the expected theoretical uncertainties, we see that the NLO
result differs much less from N-body beyond $k \simeq 0.10\ihMpc$ than expected
from the missing two-loop correction, and similarly the NNLO result remains in
good agreement above $k \simeq 0.10\ihMpc$, indicating overfitting. Ideally, we
would be able to calibrate the EFT parameters precisely below $k=0.05\ihMpc$
where the perturbative uncertainty is small, but in practice the moderate
simulation uncertainty (even after cosmic variance is reduced), a small number
of grid points and finite box size effects complicates the picture.

To gauge the extent of overfitting at NLO, we introduce a 0-parameter NLO case
where the $\gamma_1$ parameter is fixed to that in the NNLO case with the same
pivot scale. At NNLO, there is a larger window in which the theoretical
uncertainty is still small and hence more data points to measure $\gamma_1$
precisely. Moreover, in our renormalization scheme $\gamma_1$ remains the same
after adding the two-loop correction, up to numerical uncertainty and
overfitting. Therefore, the degree of overfitting is reduced in the
NLO~$\bar{\gamma}_1 = \bar{\gamma}_1^{\text{[NNLO]}}$ case.

\begin{figure}[t]
    \centering
    \includegraphics[width=0.5\linewidth]{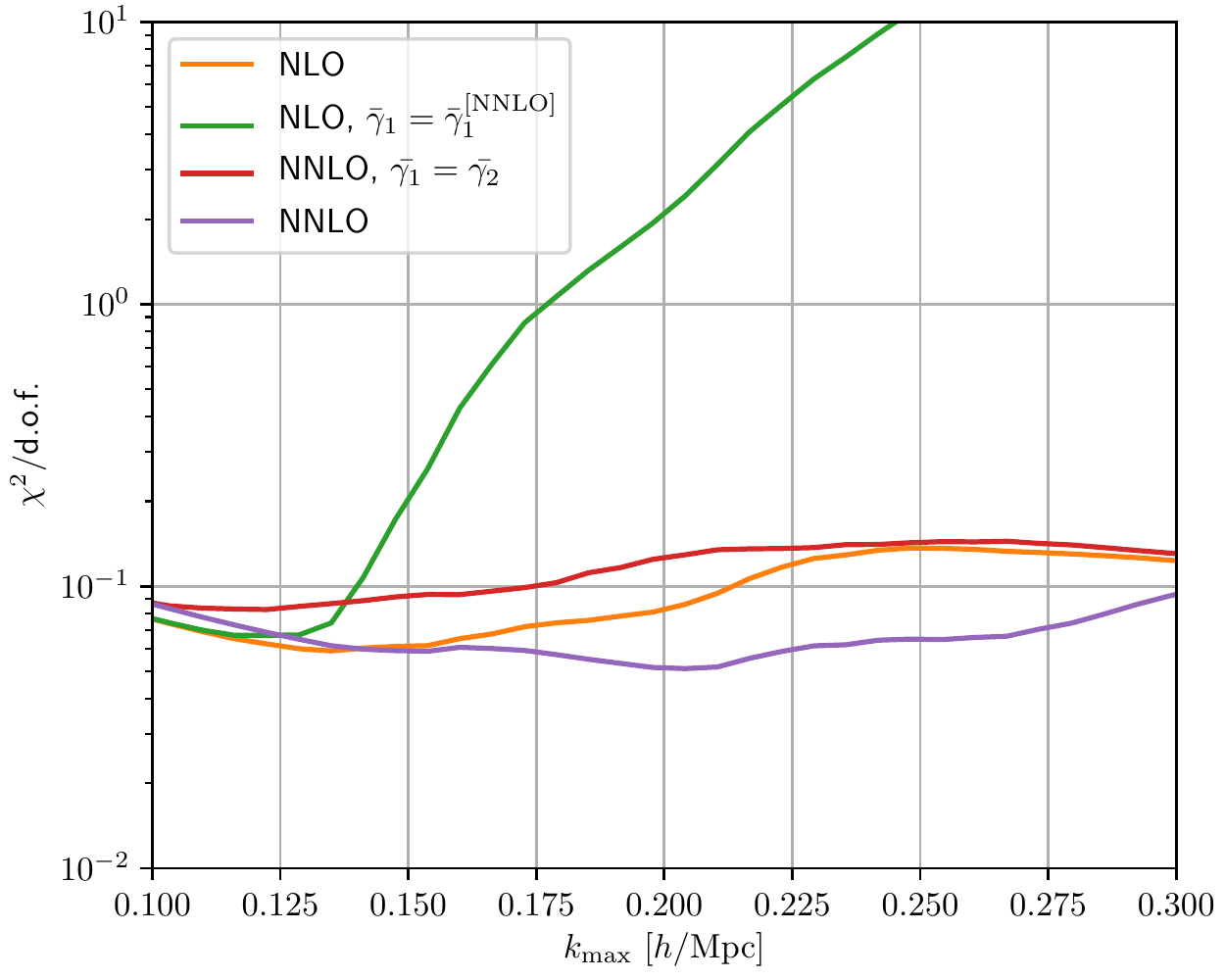}
    \caption{Reduced $\chi^2$ for the various calibration cases at NLO and NNLO
    as a function of the pivot scale $\kmax$.}
    \label{fig:chi2_order}
\end{figure}

The $\chi^2$ per degree of freedom (d.o.f.) is shown for the different fits in
Fig.~\ref{fig:chi2_order}. Since we are ultimately interested in assessing the
effect of non-linear neutrino perturbations on the power spectrum, i.e.\
comparing the performance of the 1F and 2F schemes, we chose a simple estimator
for the N-body uncertainty, for example not taking into account correlations
between different bins. This is reflected by the small
$\chi^2/\mathrm{d.o.f.} \ll 1$. We find that the $\bar{\gamma}_1 =
\bar{\gamma}_2$ ansatz at NNLO proposed by~\cite{Baldauf:2015aha} does not work
as well when including the exact time-dependence and effect of neutrinos, and
even performs worse than the NLO case at $k \simeq 0.12\ihMpc$.

\begin{figure}[t]
    \centering
    \includegraphics[width=\linewidth]{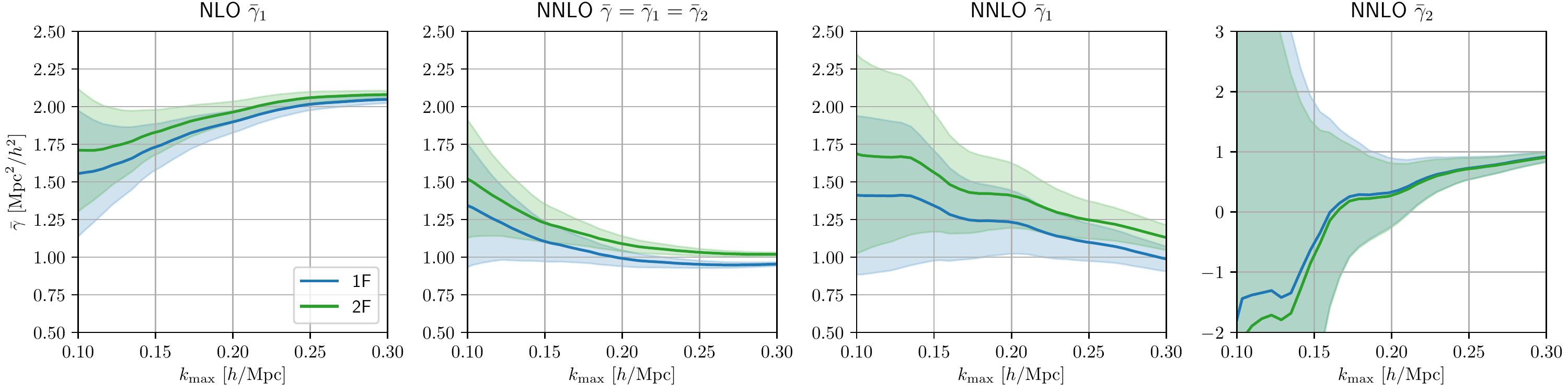}
    \caption{Measured parameters $\gamma_1$, $\gamma_2$ and
    $\gamma = \gamma_1 = \gamma_2$ as a function of $k_{\mathrm{max}}$. We show
    the results from the 1F scheme (blue) and 2F scheme (green) for a neutrino
    mass sum $\Mnu = 0.1~\eV$. The shaded regions indicate $1\sigma$
    uncertainty.}
    \label{fig:gamma_order}
\end{figure}

The best-fit EFT parameters $\bar{\gamma}_1$ and $\bar{\gamma}_2$ as a function
of $\kmax$ are displayed in Fig.~\ref{fig:gamma_order}. We show the results from
both schemes 1F and 2F, indicating $1\sigma$ uncertainty in the shaded bands.
At NLO, we find that the $\bar{\gamma}_1$ measurement is consistent with a
constant up to about $\kmax \simeq 0.11\ihMpc$ where the parameter starts
exhibiting a running. At NNLO (with both parameters varied), this plateau for
$\bar{\gamma}_1$ extends to $\kmax \simeq 0.14\ihMpc$, and the measurement
within $1\sigma$ is consistent with a constant up to around $0.2\ihMpc$. The
$\bar{\gamma}_2$ parameter (rightmost panel) can only be constrained beyond
$\kmax \simeq 0.15\ihMpc$, because the two-loop counterterm (i.e.\ the
subtracted single-hard limit~\eqref{eq:P2L_ctr}) is exceedingly small for lower
wavenumbers. We find that when both parameters are allowed to vary at NNLO,
they assume different values, giving another confirmation that the NNLO
1-parameter $\bar{\gamma}_1 = \bar{\gamma}_2$ model cannot match the data with
the same accuracy. In the second leftmost plot we show the best-fit parameter
in this model. It features a considerable running compared to the other cases
even at low $\kmax \simeq 0.1\ihMpc$, suggesting a degree of overfitting
already from large scales. Before commenting on the difference in the
calibrated EFT parameters in the 1F and 2F schemes as is seen in
Fig.~\ref{fig:gamma_order}, we discuss the power spectrum and $\chi^2$ in the
two schemes.

\begin{figure}[t]
    \centering
    \includegraphics[width=0.7\linewidth]{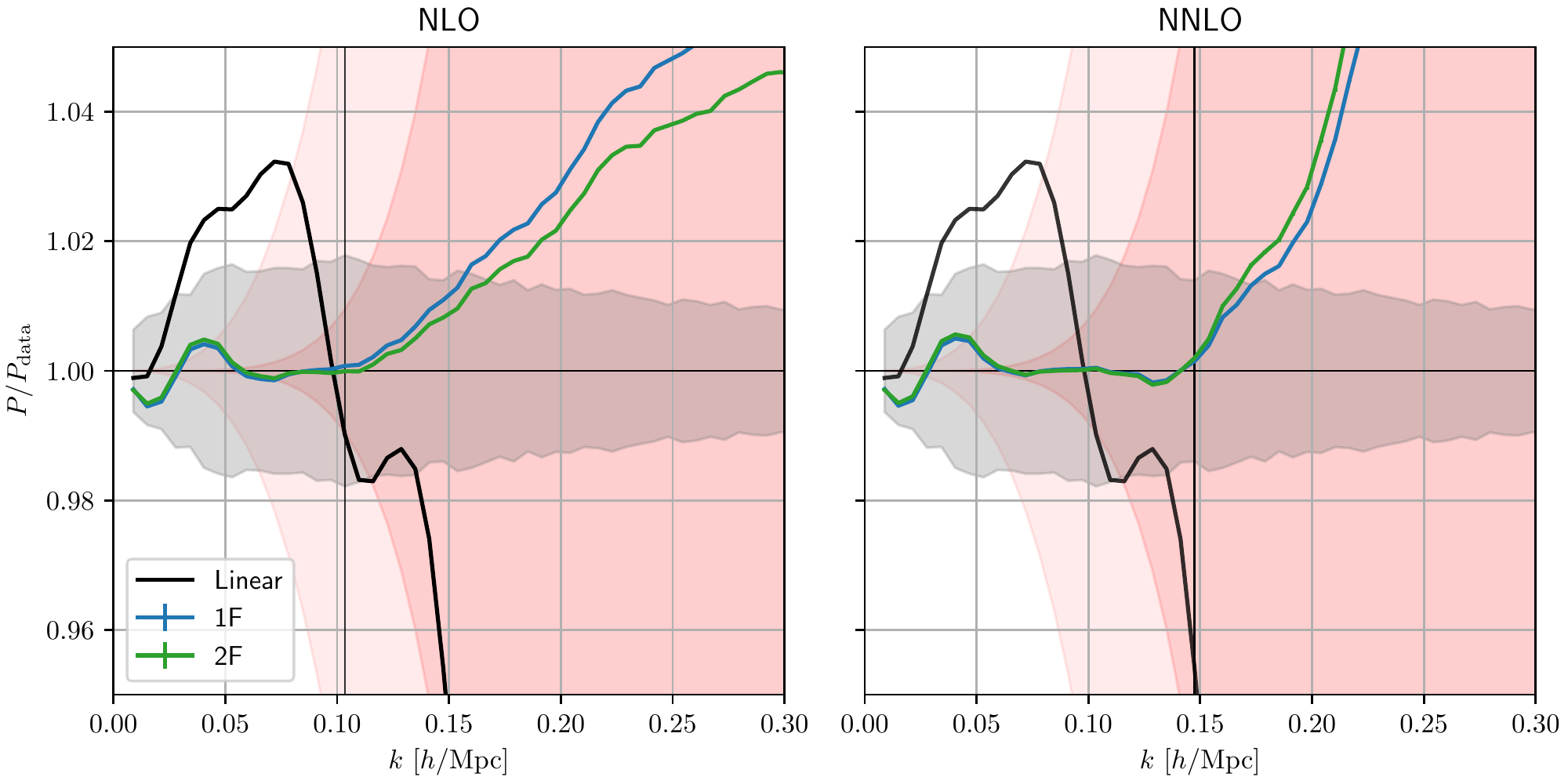}
    \caption{CDM+baryon power spectrum at NLO (left) and NNLO (right) in the
    1F (blue) and 2F (green) schemes, normalized to the N-body result. The gray
    band indicates the N-body uncertainty, while the red shaded area display
    estimated theoretical uncertainty at one-loop (light shading) and two-loop
    (darker shading). The pivot scale is $\kmax = 0.148\ihMpc$ and the sum of
    neutrino masses is $\Mnu = 0.1~\eV$.}
    \label{fig:pk_schemes}
\end{figure}

\begin{figure}[t]
    \centering
    \includegraphics[width=0.48\linewidth]{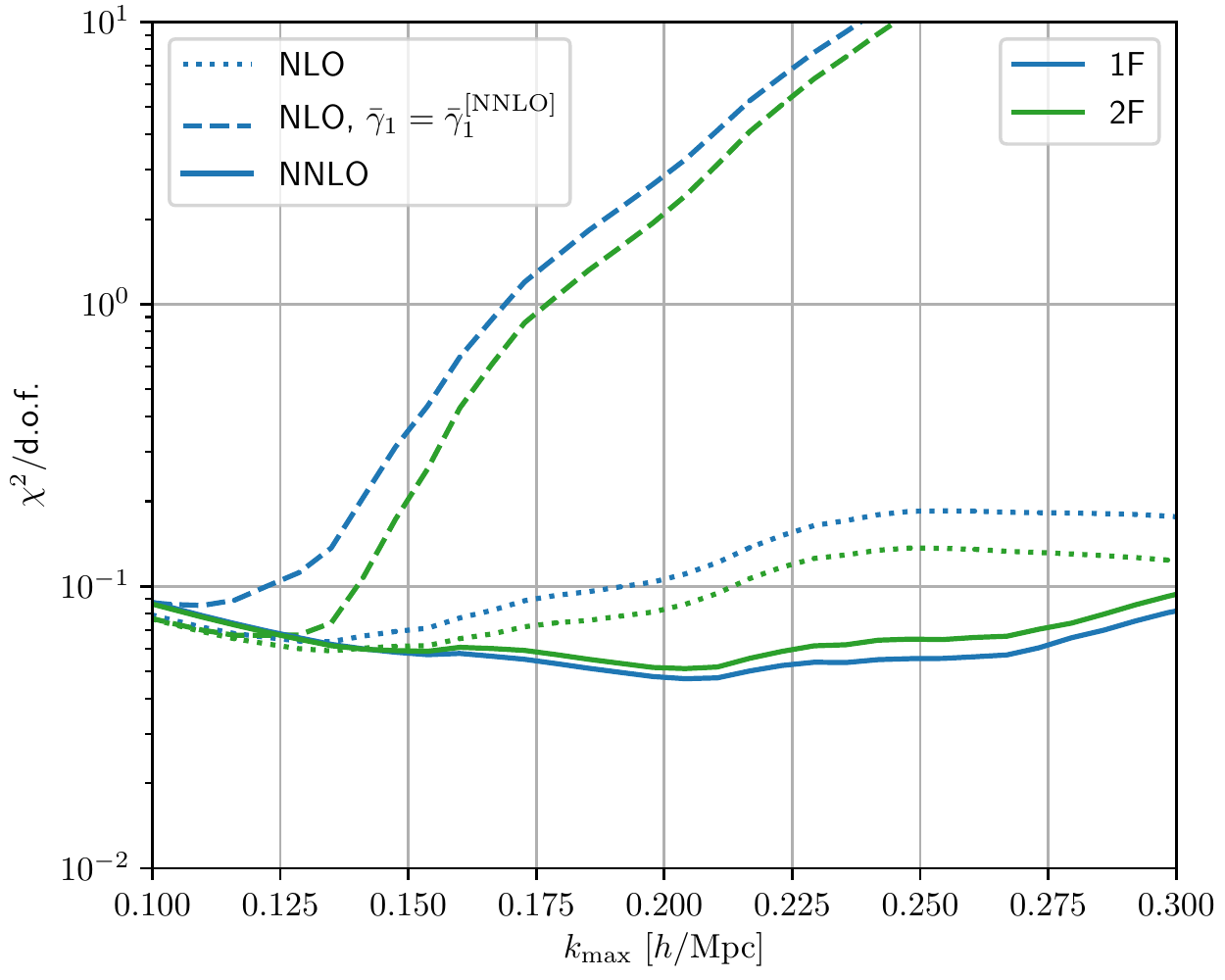}
    \includegraphics[width=0.48\linewidth]{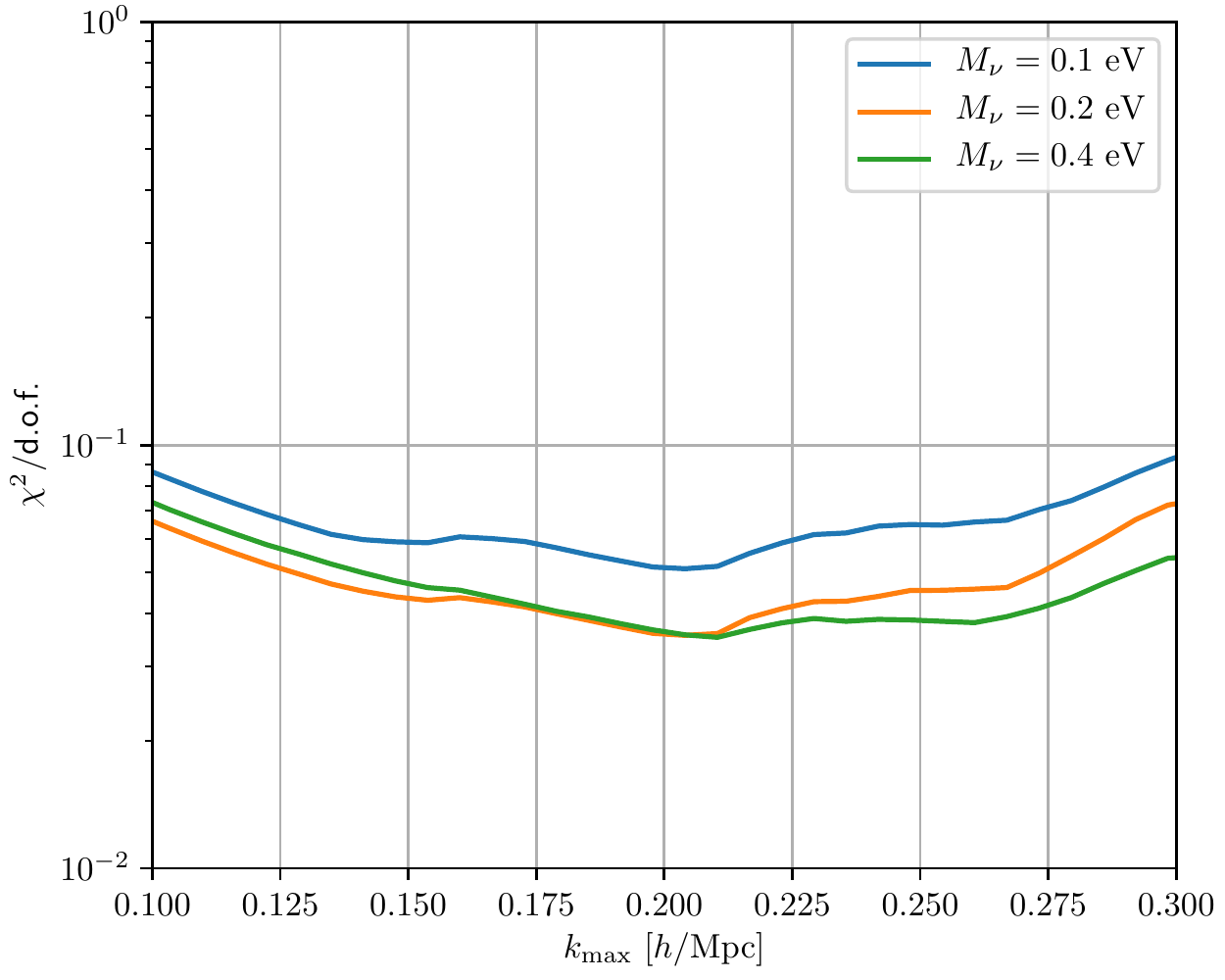}
    \caption{\emph{Left:} Reduced $\chi^2$ as a function of $\kmax$ for the 1F
    and 2F schemes, at NLO (1- and 0-parameter) and NNLO (2-parameter) in
    perturbation theory. \emph{Right:} Reduced $\chi^2$ for different neutrino
    masses, computed in the 2F scheme at NNLO (2-parameter fit).}
    \label{fig:chi2_schemes}
\end{figure}

Next, we compare the 1F and 2F schemes for $\Mnu = 0.1~\eV$. We find similar
conclusions for the other neutrino mass cosmologies. The power spectra for the
two schemes are shown in Fig.~\ref{fig:pk_schemes}, using $\kmax = 0.15\ihMpc$,
and the $\chi^2/\mathrm{d.o.f.}$ as a function of the pivot scale is shown in
the left panel of Fig.~\ref{fig:chi2_schemes}. We find that both schemes can
broadly model the data equally accurately, with some minimal differences. This
suggests that the difference between the schemes for the bare density power
spectrum (arising mostly due to relaxing the EdS approximation, see
Fig.~\ref{fig:loop_corr} and discussion in Sec.~\ref{sec:bare_pk_comparison})
can to large extent be absorbed into the counterterms. In particular, the
difference can be captured by the $k^2 P_0$-counterterm leading to a shift in
the value $\bar{\gamma}_1$. This is exactly what we see in
Fig.~\ref{fig:gamma_order}. Both in the NLO and NNLO cases, we have an
approximate shift $\Delta\bar{\gamma}_1 \approx - 0.2~\Mpc^2/h^2$ between the
1F and 2F schemes. The shift is consistent with the findings
of~\cite{Baldauf:2021zlt} for the one-loop bispectrum with the EdS
approximation relaxed. Furthermore, we find that the two-loop EFT parameter
$\bar{\gamma}_2$ is unchanged in the 2F scheme as compared to 1F, as seen for
$\kmax > 0.15\ihMpc$ in the rightmost panel of Fig.~\ref{fig:gamma_order}.

Finally, we consider the three cosmologies with $\Mnu = 0.1$, $0.2$ and
$0.4~\eV$, computed at NNLO (2-parameter) in the full 2F solution. The reduced
$\chi^2$ for the different models is displayed in the right panel of
Fig.~\ref{fig:chi2_schemes}. By accident, the ``bump'' feature from finite box
effects is slightly smaller for $\Mnu = 0.2~\eV$ than the other neutrino
masses, leading to a smaller $\chi^2$ for low $\kmax$. Moreover, for $\Mnu =
0.4~\eV$, the linear 2F solution is off by almost a percent for $k<0.01\ihMpc$
(see Fig.~\ref{fig:lin_comparison}), yielding a slightly larger $\chi^2$ around
$\kmin$. Apart from these insignificant discrepancies, we find that all
neutrino models can match the N-body results accurately using the 2F scheme.
This is also seen from the power spectra plots in Fig.~\ref{fig:pk_mass}.

\begin{figure}[t]
    \centering
    \includegraphics[width=\linewidth]{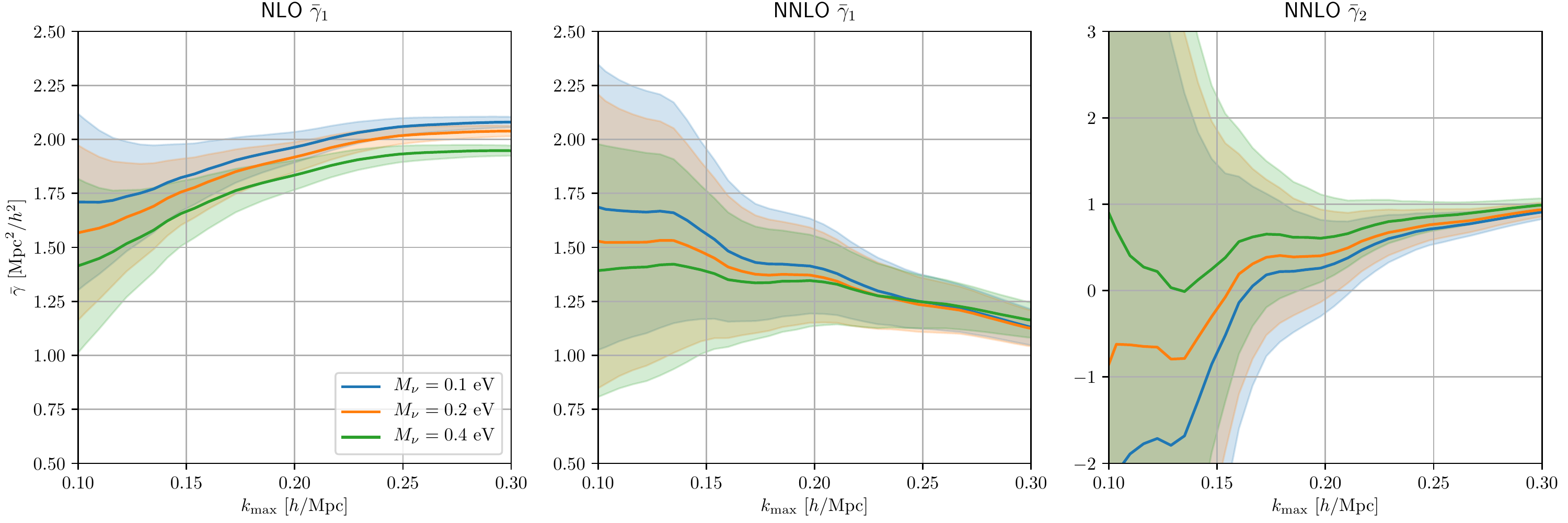}
    \caption{Measured parameters $\bar{\gamma}_1$, $\bar{\gamma}_2$ and as a
    function of $\kmax$ for the different cosmologies with $\Mnu = 0.1$, $0.2$
    and $0.4~\eV$. The shaded regions indicate $1\sigma$ uncertainty.}
    \label{fig:gamma_mass}
\end{figure}

The best-fit EFT parameters for the different neutrino mass models are shown in
Fig.~\ref{fig:gamma_mass}. In the left plot we use the 1-parameter NLO approach
while in the middle and right plots we show the NNLO 2-parameter model. It is
not surprising that the parameters acquire different values for the different
cosmologies: the part of the counterterms correcting for the inaccuracy of SPT is
altered due to different values of the displacement dispersion $\sigmad$ and the
$c$-coefficient, and the part accounting for the impact of actual short-scale
physics changes due to the impact of massive neutrinos on strongly non-linear scales.
At NNLO, we measure a shift $\Delta \bar{\gamma}_1 \simeq -0.14~\Mpc^2/h^2$ at $\kmax = 0.14\ihMpc$ and
$\Delta \bar{\gamma}_2 \simeq -0.2~\Mpc^2/h^2$ at $\kmax = 0.17\ihMpc$ when
doubling the sum of neutrino masses. We quote the value of the $c$-coefficient
as well as the measured EFT parameters $\gamma_1(\Lambda)$ and
$\gamma_2(\Lambda)$ for $\Lambda = 1\ihMpc$ and $\Lambda \to\infty$ for the
different neutrino cosmologies in Tab.~\ref{tab:EFT_params}.

\begin{table*}[t]
    \centering
    \caption{Measured EFT parameters and constant $c$ for different neutrino
    masses with cutoff $\Lambda = 1\ihMpc$ and pivot scale
    $\kpivot = 0.15\ihMpc$.}
    \label{tab:EFT_params}
    {\def\arraystretch{1.55}
    \begin{tabular}{l|rrrrr}
        \hline
                            & $c$       & $\gamma_1(\Lambda)$ & $\bar{\gamma}_1$  & $\gamma_2(\Lambda)$ & $\bar{\gamma_2}$ \\
        \hline
        $\Mnu = 0.1~\eV$    & $ 0.269 $ & $ 1.81 \pm 0.41 $   & $ 1.58 \pm 0.41 $ & $ -0.62  \pm 2.4 $  & $ -0.85 \pm 2.4 $ \\
        $\Mnu = 0.2~\eV$    & $ 0.254 $ & $ 1.71 \pm 0.42 $   & $ 1.48 \pm 0.42 $ & $ -0.072 \pm 2.3 $  & $ -0.3 \pm 2.3  $ \\
        $\Mnu = 0.4~\eV$    & $ 0.228 $ & $ 1.61 \pm 0.39 $   & $ 1.4  \pm 0.39 $ & $ 0.46 \pm 2.1   $  & $ 0.25 \pm 2.1  $ \\
    \end{tabular}
    }
\end{table*}

In summary, we can accurately match the Quijote N-body data both at NLO and
NNLO using the full 2F solution, however being subject to a certain degree of
overfitting. The window where the EFT parameters can be precisely measured is
relatively small and involve uncertainties from the N-body simulation. We
compared the 1F and 2F schemes and found that the difference of the
unrenormalized power spectrum can largely be absorbed in the one-loop
counterterm, leading to a shift in the EFT parameter $\bar{\gamma}_1$.

\section{Conclusions}
\label{sec:conclusions}

In this work we have performed a precision calculation of the CDM+baryon density and velocity power
spectrum at next-to-next-to-leading (two-loop) order taking the full time- and scale-dependence of neutrino perturbations beyond
the linear level as well as exact $\Lambda$CDM ($+\Mnu$) time-dependence into
account. Our aim is to assess the accuracy of the commonly used approach in
which the aforementioned effects are neglected.

To describe the impact of neutrinos in gravitational collapse, we use a hybrid
two-component fluid scheme introduced in~\cite{Blas:2014} and further developed
in~\cite{Garny:2020ilv}. The idea is that neutrinos can only be described by a
fluid well after they have become non-relativistic, because the lower and
higher order multipoles effectively decouple by powers of $T_{\nu}/m_{\nu}$.
Therefore, we can use the full, linear Boltzmann hierarchy until an
intermediate redshift $z = 25$, where we map the equations onto a two-component
fluid consisting of CDM+baryons (one joint component) and neutrinos. The large
neutrino free-streaming is reflected in an effective neutrino sound velocity,
which we compute from linear theory. The two-component fluid system can readily
be realized in the framework for computing loop corrections in cosmologies with
general time- and scale-dependence introduced in~\cite{Garny:2020ilv}.

We compare the (unrenormalized) two-loop CDM+baryon density and velocity power
spectra at $z = 0$ from the full solution including neutrino perturbations
beyond the linear level and exact time-dependence (2F scheme) to the commonly used simplified
treatment with only linear neutrino perturbations and EdS dynamics for CDM+baryons (1F scheme).
For the density spectrum, the main difference arises due to the departure from
EdS, leading to 0.7\% deviation at $k = 0.15\ihMpc$ (consistent with the
results in~\cite{Garny:2020ilv}), approximately neutrino mass independent. On
the other hand, for the velocity spectrum, the deviation in the 1F scheme is
neutrino mass dependent; for the largest neutrino mass $\Mnu = 0.4~\eV$ the
deviation is larger than a percent for $k\gtrsim 0.08\ihMpc$. For $k = 0.15\ihMpc$
the deviation in the velocity or cross power spectra up to two-loop
are $2.7\%, 1.3\%, 0.8\%$ for $M_\nu=0.4, 0.2, 0.1~\eV$, respectively, and
even somewhat larger, $2.7\%, 2.0\%, 1.6\%$, when going only up to one-loop, due to
a partial cancellation between one- and two-loop terms. Therefore, the commonly
used 1F approximation is questionable for the velocity spectra when aiming at percent
accuracy, implying that the 2F scheme should be used to access the neutrino mass
sensitivity in redshift space.

Furthermore, we
check to what extent the two-loop power spectrum on mildly non-linear scales in
the presence of free-streaming massive neutrinos can be reproduced by a
massless model by adjusting the overall amplitude of fluctuations. We find that
the massless model can reproduce the total matter density power spectrum of the smallest neutrino
mass to within a percent, but for the larger neutrino masses the neutrino
free-streaming scale $\kFS$ is larger and mimicking the power spectrum in the
massless model is more challenging.

To assess the performance of the full 2F solution and the simpler 1F scheme as
accurately as possible, and to compare to N-body data, we take into account EFT
corrections as well as effects of large bulk flows from the IR. We find that
the $k^2$ scaling of the kernels -- expected from momentum conservation in the
limit when the external wavenumber $k$ goes to zero (or the other momenta go to
infinity) -- is slightly spoiled by the free-streaming scale, but recovered when
$k\ll\kFS$ or $k\gg\kFS$. Therefore, we assume a large scale separation $\kFS
\ll k \ll q$, where $q$ is the scale of hard loop momenta, when analysing the
UV limit of the 2F model. This assumption is appropriate for the smallest neutrino
masses, but breaks down for the largest neutrino mass. At one-loop, we show
that the usual $k^2 P_0(k)$ counterterm absorbs the cutoff-dependence. The
double-hard region of the two-loop contribution can also be corrected by this
counterterm. For the single-hard region, we employ a numerical treatment
yielding a counterterm that exactly captures the scale-dependence of the
UV-limit of SPT, with one extra EFT parameter.

We compare the perturbative results of both the 1F and 2F schemes to N-body
data, at different orders in perturbation theory and with various number of
free parameters. The power spectrum in the 2F scheme matches the N-body data
within a percent up to $k\simeq 0.12\ihMpc$ and $k\simeq 0.16\ihMpc$ for NLO
and NNLO, respectively, using a pivot scale $\kmax = 0.148\ihMpc$. We find that
the 1F scheme can achieve similar accuracy, indicating that the deviation in
the bare power spectrum can be absorbed by the counterterms. In particular, the
main source of error in the 1F density power spectrum comes from the EdS approximation,
and it is largely degenerate with the one-loop counterterm. We measure a shift
in the EFT parameter $\Delta\bar{\gamma}_1 = - 0.2~\Mpc^2/h^2$ between the 1F
and 2F schemes that accounts for the discrepancy.

In total, we have scrutinized the effect of massive neutrino perturbations
taking into account the full impact of time- and scale-dependent growth on non-linear kernels for CDM+baryon as well as neutrino density and velocity.
We further capture the exact time-dependence of $\Lambda$CDM ($+\Mnu$). The impact on the two-loop density
power spectrum is to great extent degenerate with counterterms. The influence of
time- and scale-dependent growth due to massive neutrinos is larger on the velocity spectrum, suggesting that the full 2F treatment for neutrinos
is warranted to access the neutrino mass information encoded in redshift space distortions. This motivates an analysis of the power spectrum
in redshift space within the 2F scheme, which we leave to future work.

\acknowledgments{%
We thank Francisco Villaescusa-Navarro for helpful comments on the Quijote simulation suite.
This work was supported by the DFG Collaborative Research Institution
\emph{Neutrinos and Dark Matter in Astro- and Particle Physics}
(\href{www.sfb1258.de}{SFB 1258}).
}

% \bibliographystyle{JHEP}
% \bibliography{main}

\providecommand{\href}[2]{#2}\begingroup\raggedright\endgroup

\end{document}